\documentclass[aps,prd,showpacs,preprintnumbers]{revtex4}
\usepackage{graphicx}                                 
\usepackage{amsmath,amsfonts}
\graphicspath{{./Figures/}}                           

\usepackage{epsfig}
\usepackage{amssymb}
\usepackage{amsfonts}
\usepackage{float}
\newcommand{\beq}{\begin{equation}}
\newcommand{\eeq}{\end{equation}}
\newcommand{\bea}{\begin{eqnarray}}
\newcommand{\eea}{\end{eqnarray}}
\newcommand{\beas}{\begin{eqnarray*}}
\newcommand{\eeas}{\end{eqnarray*}}
\newcommand{\eq}{\begin{equation}}
\newcommand{\en}{\end{equation}}
\newcommand{\eqa}{\begin{eqnarray}}
\newcommand{\ena}{\end{eqnarray}}

\begin{document}

\preprint{ITEP-LAT/2008-05, HU-EP-08/19}

\title{The topological structure of $SU(2)$ gluodynamics at $T > 0$ : \\
an analysis using the Symanzik action and Neuberger overlap fermions}
\author{V.~G.~Bornyakov}
\affiliation{Institute for High Energy Physics, Protvino, 142281, Russia}
\affiliation{Institute of Theoretical and  Experimental Physics,
B.~Cheremushkinskaya~25, Moscow, 117259, Russia}
\author{E.~V.~Luschevskaya}
\affiliation{Institute of Theoretical and  Experimental Physics,
B.~Cheremushkinskaya~25, Moscow, 117259, Russia}
\author{S.~M.~Morozov}
\affiliation{Institute of Theoretical and  Experimental Physics,
B.~Cheremushkinskaya~25, Moscow, 117259, Russia}
\author{M.~I.~Polikarpov}
\affiliation{Institute of Theoretical and  Experimental Physics,
B.~Cheremushkinskaya~25, Moscow, 117259, Russia}
\author{E.-M.~Ilgenfritz}
\affiliation{Institut f\"ur Physik, Karl-Franzens-Universit\"at 
Graz, Universit\"atsplatz 5, A-8010 Graz, Austria}
\affiliation{Humboldt-Universit\"at zu Berlin, Institut f\"ur Physik,
Newtonstrasse~15, 12489 Berlin, Germany}
\author{M.~M\"uller-Preussker}
\affiliation{Humboldt-Universit\"at zu Berlin, Institut f\"ur Physik,
Newtonstrasse~15, 12489 Berlin, Germany}

\date{February, 22, 2009}

\begin{abstract}
We study $SU(2)$ gluodynamics at finite temperature
on both sides of the deconfining phase transition.
We create the lattice ensembles using the tree-level
tadpole-improved Symanzik action.
The Neuberger overlap Dirac operator is used to determine
the following three aspects of vacuum structure:
({\it i}) The topological susceptibility
is evaluated at various temperatures across the phase
transition,
({\it ii}) the overlap fermion spectral density is
determined and found to depend on the Polyakov loop
above the phase transition and
({\it iii}) the corresponding localization properties
of low-lying eigenmodes are investigated.
Finally, we compare with zero temperature results.
\end{abstract}


\maketitle

\section{Introduction}
\label{sec:introduction}

More than ten years ago, in a model generalizing random matrix theory,
M. A. Stephanov~\cite{Stephanov} has built in the feature
that in $SU(3)$ gluodynamics
above $T_c$ the different Polyakov loop sectors should behave differently.
In the sectors with complex-valued average Polyakov loop the chiral
condensate was expected to turn to zero at a temperature $T$ substantially
higher than $T_c$, the deconfinement temperature.
Prior to this paper, similar observations
in quenched $SU(3)$ lattice simulations had been reported by Chandrasekharan
and Christ~\cite{Chandrasekharan1,Chandrasekharan2}.
For $SU(2)$ lattice gluodynamics, where the Polyakov loop is real, it was
predicted that the chiral condensate stays non-zero,
$<\bar{\psi}\psi> \neq 0$,
for all temperatures $T > T_c$ in the sector with a negative average
Polyakov loop, $L<0$.

Simulating quenched $SU(3)$, Gattringer {\it et al.}~\cite{Gattringer}
came to a different result. These authors used the spectral gap in the
spectrum of the Dirac operator as an order parameter for the 
restoration of chiral symmetry. It was found that the spectral gap
opens up at one single temperature $T=T_c$ in all three $Z_{3}$ sectors. 

In $SU(2)$ gluodynamics this question has not been clarified before.
In this case the behavior may be completely different. Indeed, there 
are pure vacuum gauge field configurations with $L=1$ for which the 
Dirac operator has trivial zero modes for periodic fermionic boundary
conditions. Contrary to $SU(3)$, for $SU(2)$ (as well as for any even
number of colors) through a non-periodic $Z(2)$ gauge transformation
with $- {\bf I} \in Z(2)$ this is equivalent to the existence of zero 
modes for antiperiodic fermion boundary conditions in the case of $L=-1$.
In the light of this argument, we shall examine whether Stephanov's 
prediction for the Dirac spectrum remains valid in the case of $SU(2)$ 
gluodynamics in the deconfined phase. Preliminary results were presented 
in~\cite{Lattice2007}.

We base our study on Neuberger's overlap Dirac operator~\cite{Neuberger}.
Our interest will be concentrated on its spectrum, how it is related to
the average Polyakov loop, and how this is reflected by the localization
properties of the modes in various parts of the spectrum.
Because of the index theorem~\cite{Laliena,Adams}, the zero modes give us
easy access to the topological charge $Q$ of the lattice configurations,
which is sufficient to calculate the topological susceptibility.
In our present work, in contrast to Ref.~\cite{calorons}, we will not
exploit the capability of the overlap Dirac operator to provide us with
a tool to analyse the topological charge density~\cite{Niedermayer}.

A first study of the overlap Dirac operator $D_{ov}$
in $SU(2)$ gluodynamics at finite temperature has been undertaken by the
authors of  Ref.~\cite{Edwards}. They have used the usual Wilson
gauge field action which means that for the same lattice spacing their
configurations are substantially more rough than ours. We compare our
results with theirs at several places in the following.

The outline of the paper is as follows.
In Sect.~\ref{sec:action} we explain
the action used and quote the ensembles investigated.
In Sect.~\ref{sec:overlap_operator} we discuss the structure and handling
of Neuberger's overlap operator. In Sect.~\ref{sec:topological_susceptibility}
we report about the $T$-dependence of the topological susceptibility and
in Sect.~\ref{sec:spectral_density} about the temperature {\it and} Polyakov
loop dependence of the spectral density. In Sect.~\ref{sec:spectral_gap} we
define the spectral gap as a deconfinement order parameter.
In Sect.~\ref{sec:localization} the localization properties of the
Dirac eigenfunctions are described in the confined and deconfined phase,
and the dependence on the Polyakov loop is pointed out for the latter.
In Sect.~\ref{sec:volume_and_discretization} we assess the volume and
lattice-spacing dependence of all discussed observables. In a separate
Sect.~\ref{sec:Edwards} we consider an exceptional, relatively well
separated part of the non-zero modes with very small eigenvalues. After a
discussion of topological susceptibility, spectral density and localization
for the case of zero temperature in Sect.~\ref{sec:zero_temperature}, a
summary is given in Sect.~\ref{sec:summary}.

\section{The improved action}
\label{sec:action}

In an initial study, partly reported at Lattice 2007~\cite{Lattice2007},
we have analyzed ensembles of $O(100)$ statistically independent
$SU(2)$ configurations,
generated in the quenched approximation with the tadpole improved
Symanzik action on lattices of size $20^3\times6$.
This action is known to suppress
dislocations, that would lead to unphysical near-to-zero modes of the
Wilson-Dirac operator.
These are highly unwanted because they lead to difficulties to choose
a uniform $\rho$ parameter (see below).
The form of the action is :
\begin{equation}
S=\beta_{\rm imp}\sum_{pl}S_{pl}-\frac{\beta_{\rm imp}}{20u_{0}^2}\sum_{rt}S_{rt} \; ,
\label{eq:action}
\end{equation}
where $S_{pl}$ and $S_{rt}$ denote the plaquette and $1\times2$
rectangular loop
terms in the action, $S_{pl,rt}= (1/2)~{\rm Tr}~(1-U_{pl,rt})$. The factor
$u_{0}=(W_{1\times1})^{1/4}$ is the {\it input} tadpole factor.
It is selfconsistently
determined from $W_{1\times1}=\langle (1/2)~{\rm Tr}~U_{pl} \rangle$
computed at zero temperature~\cite{universality}.
The deconfining phase transition, which is of second order for $SU(2)$
gluodynamics, is not easy to locate precisely~\cite{Engels} because of
strong finite volume effects.
We have approximately determined the critical $\beta_{{\rm imp},c}=3.248(2)$
on lattices of size $20^3\times6$, {\it i.e.} for $N_{\tau}=6$.
This corresponds to a ratio $T_{c}/\sqrt{\sigma}=0.71(2)$~\cite{calorons}.

To check for finite spatial volume effects, we made additional simulations
also on $24^3\times6$ lattices as well as on smaller lattices $16^3\times6$
and $12^3\times6$.
Furthermore, we also checked whether our results are close to the continuum
limit. For this check we have simulated the neighbourhood of the transition
on a finer lattice, with $N_{\tau}=8$, for which a critical
$\beta_{{\rm imp},c}=3.425(5)$ was found in simulations on lattices with a
spatial extent $N_s=24$. For these $24^3\times8$ lattices our actual
measurements were performed only at one temperature very close to $T_c$ in
the confinement phase.

The parameters of the lattices used in this study and the respective statistics
are reported in Table~\ref{tbl:sim_det}. The lattice scale $a$ needed to compute
$T/T_c$ has been obtained by interpolating the results of Ref.~\cite{universality}.

\begin{table}[t]
\begin{center}
\begin{tabular}{|c|c|c|c|c|}
\hline
$N_{\tau} \times N_s^3 $ & $ \beta_{imp}$ & $ T/T_c$ & $ \# conf. $ & $N_{modes}$  \\
\hline
$6 \times 12^3$          &      3.20       & 0.91    & 998, 910   &    20       \\
$6 \times 12^3$          &      3.23       & 0.97    & 893, 900   &    20       \\
\hline
$6 \times 16^3$          &      3.20       & 0.91    & 482, 517   &    30       \\
$6 \times 16^3$          &      3.23       & 0.97    & 246, 262   &    50       \\
$6 \times 16^3$          &      3.325      & 1.15    & 183, 215   &    30       \\
$6 \times 16^3$          &      3.50       & 1.55    & 308, 96    &    50       \\
\hline
$6 \times 20^3$          &      3.20       & 0.91    & 48, 49     &    50       \\
$6 \times 20^3$          &      3.23       & 0.97    & 152, 147   &    50       \\
$6 \times 20^3$          &      3.275      & 1.05    & 89, 102    &    50       \\
$6 \times 20^3$          &      3.30       & 1.10    & 42, 56    &    50       \\
$6 \times 20^3$          &      3.325      & 1.15    & 101, 98    &    50       \\
$6 \times 20^3$          &      3.35       & 1.20    & 123, 75    &    50       \\
$6 \times 20^3$          &      3.40       & 1.31    & 148, 0     &    50       \\
$6 \times 20^3$          &      3.50       & 1.55    & 89, 98     &    50       \\
$6 \times 20^3$          &      3.64       & 2.00    & 99, 98     &    50       \\
 \hline
$6 \times 24^3$          &      3.23       & 0.97    & 55, 41     &    45       \\
$6 \times 24^3$          &      3.325      & 1.15    & 115, 50    &    45       \\
$6 \times 24^3$          &      3.35       & 1.20    & 46, 00     &    45       \\
$6 \times 24^3$          &      3.50       & 1.55    & 78, 81      &    45       \\
 \hline
$8 \times 24^3$          &      3.42       & 0.99    & 28, 31     &    25       \\
\hline
$10 \times 10^3$         &      3.096      & 0     &  200   &    40       \\
$12 \times 12^3$         &      3.196      & 0     &  100   &    40       \\
$14 \times 14^3$         &      3.281      & 0     &  100   &    40       \\
$16 \times 16^3$         &      3.3555     & 0     &  100   &    40       \\
$20 \times 20^3$         &      3.5        & 0     &   99   &    20       \\
\hline
\end{tabular}
\caption{The simulations details.
The fourth column shows numbers of configurations with $L>$ and $L<0$ 
separately where the two states can be distinguished. }
\label{tbl:sim_det}
\end{center}
\end{table}

For a comparison with zero temperature, we have made simulations on symmetric lattices. Respective parameters are also given in  Table~\ref{tbl:sim_det}.
At zero temperature the inverse coupling was chosen according to the lattice size such that the physical size was kept approximately fixed and equal to $(1.4~\mathrm{fm})^4$.

\section{The massless overlap Dirac operator}
\label{sec:overlap_operator}

The massless overlap Dirac operator has the form~\cite{Neuberger}
\begin{equation}
D_{ov}=\frac{\rho}{a} \left( 1 + D_W \Big/ \sqrt{D^{\dagger}_W D_W} \right) \; ,
\label{eq:overlap}
\end{equation}
where $D_W = M - \rho/a$ is the Wilson-Dirac operator with a negative mass term
$\rho/a$, $M$ is the Wilson hopping term with $r=1$, and
$a$ is the lattice spacing.
With respect to the locality property of the overlap operator the optimal 
value of the $\rho$ parameter is found to be $\rho=1.4$ also for the
lattice ensembles under investigation. Standard boundary conditions,
anti-periodic in time and periodic in spatial directions, are imposed
to the fermionic field and the overlap Dirac operator.

In order to compute the sign function in
\begin{equation}
D_W \Big/ \sqrt{D^{\dagger}_W D_W} = \gamma_5~{\rm sign}(H_W) \; ,
\label{eq:signfunction}
\end{equation}
where $H_W = \gamma_5~D_W$ is the {\it hermitian} Wilson-Dirac operator,
we have used the minmax polynomial approximation. More precisely,
we have treated
20 to 50 lowest Wilson-Dirac eigenmodes
explicitely and used the polynomial approximation in the subspace
orthogonal to these lowest Wilson-Dirac eigenmodes.
The overlap Dirac operator $D_{ov}$ constructed this way preserves
the chiral symmetry even on a lattice with finite $a$ and allows to
exploit the perfect chiral properties of the emerging Dirac overlap
eigenmodes.
The overlap operator will be called $D$ in the following and replaces the continuum
Dirac operator $D=D_{\mu}~\gamma_{\mu}$, where $D_\mu=\partial_\mu-igA_\mu$ is
the covariant partial derivative with the gauge field background $A_{\mu}$.

\section{The topological susceptibility $\chi_{top}(T)$}
\label{sec:topological_susceptibility}

We solved the Dirac equation numerically finding its eigensystem
\begin{equation}
D~\psi_n = \lambda_n~\psi_n \, ,
\label{eq:eigensystem}
\end{equation}
restricted to the 50 lowest eigenvectors by means of ARPACK~\cite{arpack}.
They are perfectly located on the Ginsparg-Wilson circle.
By stereographic projection they are mapped to the imaginary axis. These
purely imaginary eigenvalues $\lambda_{\rm impr}$ are then representing the
$O(a)$ improved overlap Dirac operator. The spectral densities presented later
are distributions with respect to the imaginary part of $\lambda_{\rm impr}$.

\begin{figure}
\begin{tabular}{cc}
\includegraphics[scale=0.42,angle=270]{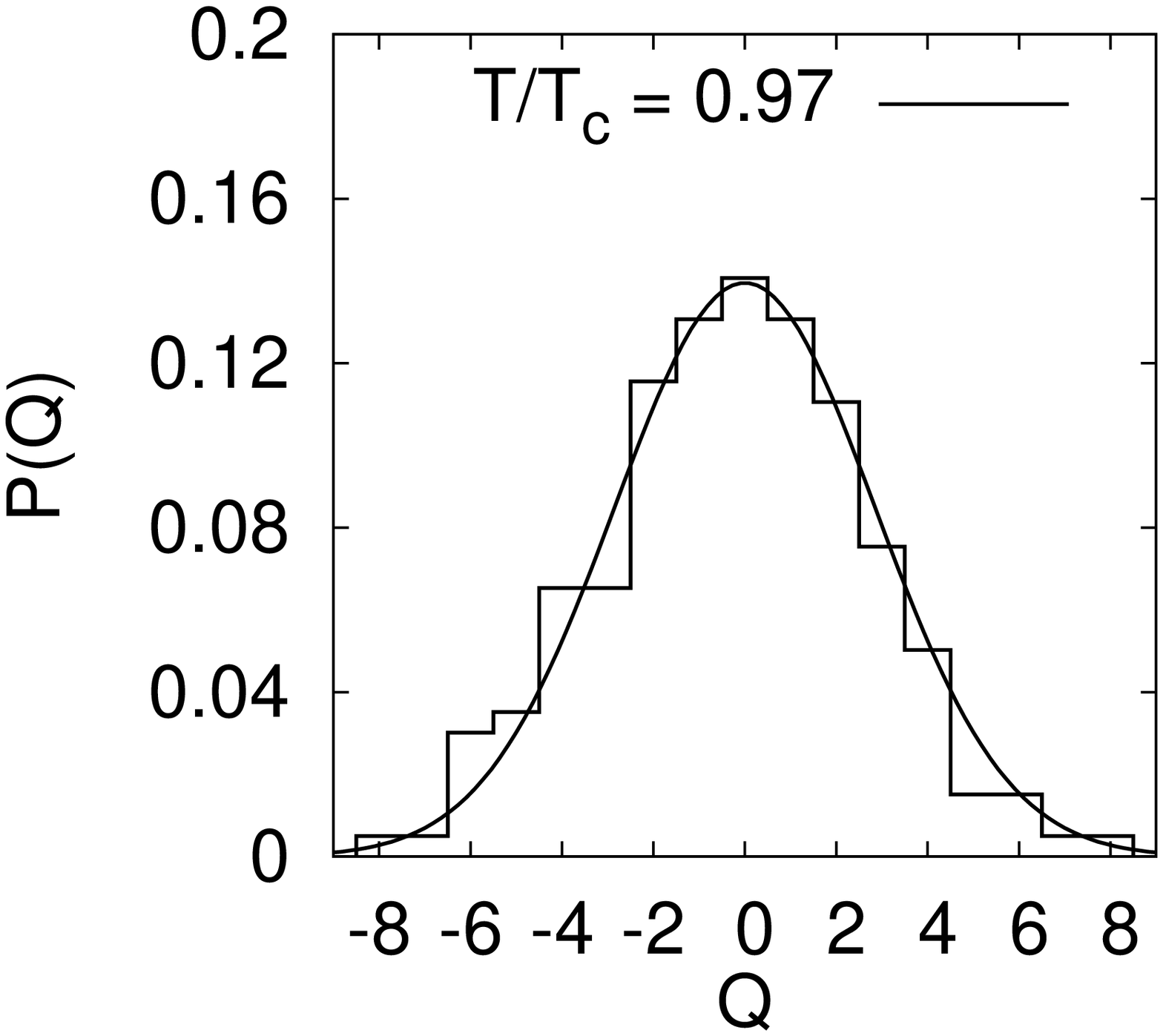} &
\includegraphics[scale=0.37,angle=270]{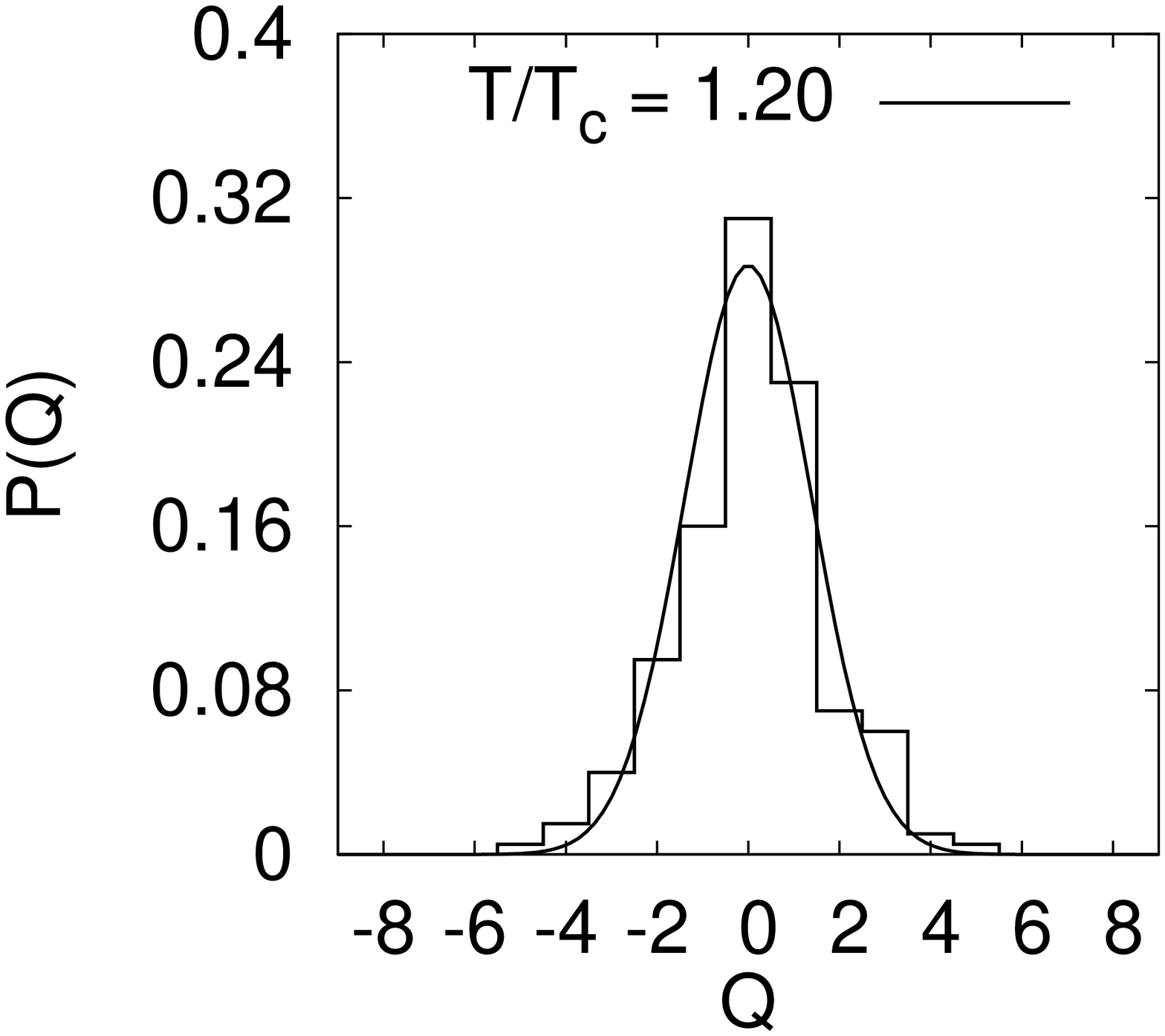} \\
\end{tabular}
\caption{Probability distributions of the topological charge $Q$ for
two temperatures below and above $T_c$ on the $20^3 \times 6$ lattice.}
\label{fig:fig1}
\end{figure}

\begin{figure}
\begin{tabular}{cc}
\includegraphics[scale=0.31,angle=270]{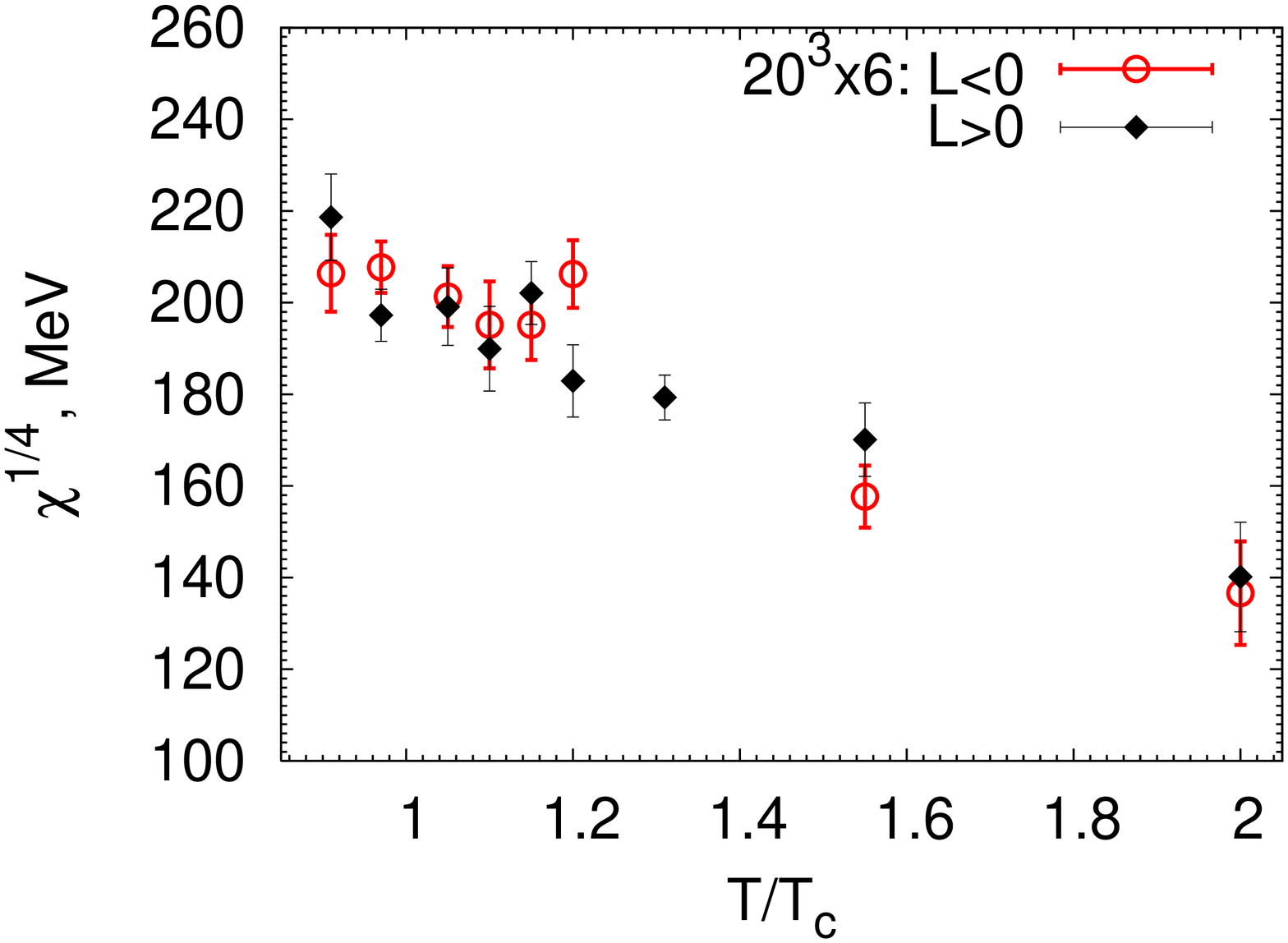} &
\includegraphics[scale=0.31,angle=270]{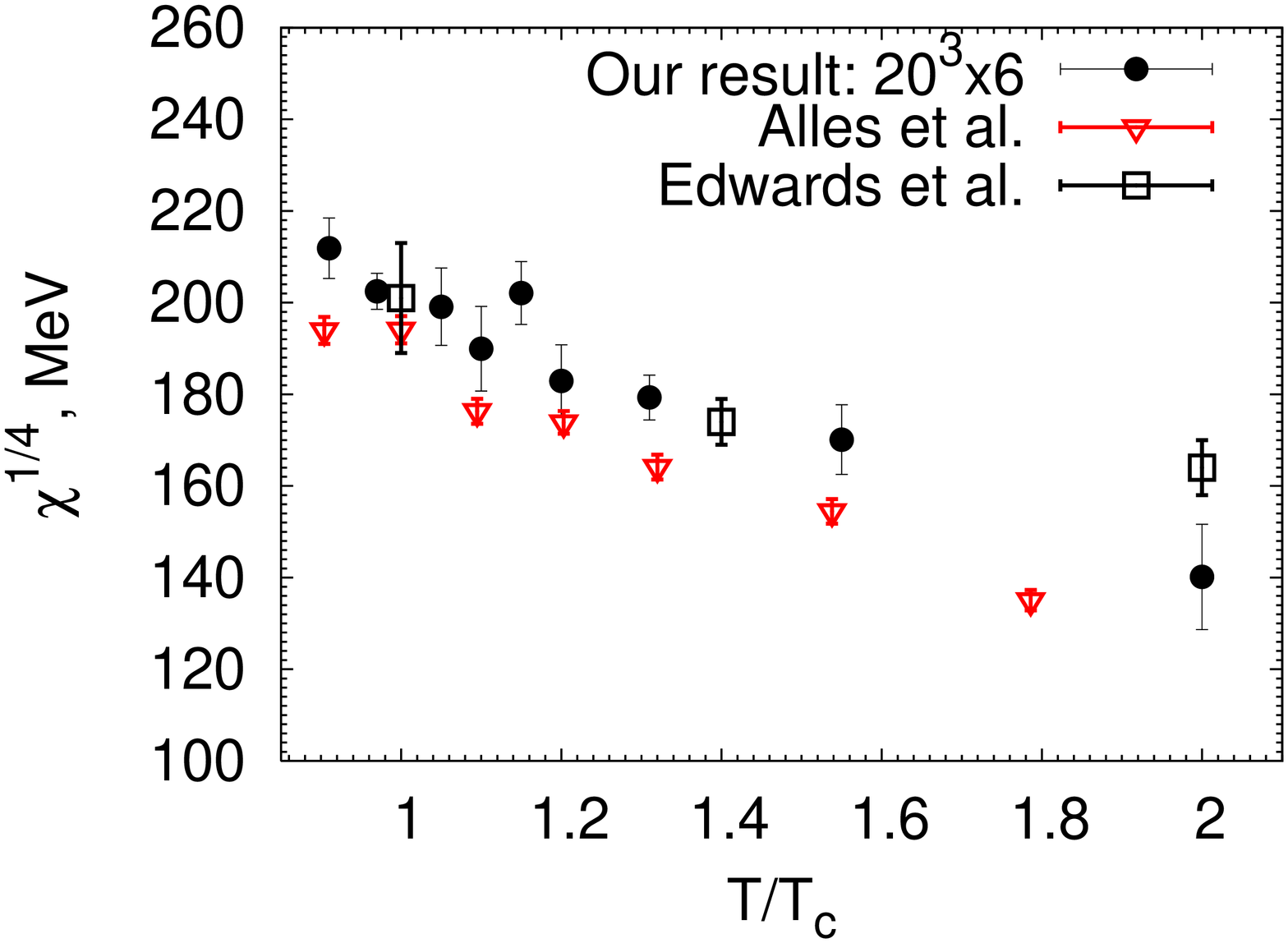} \\
\end{tabular}
\caption{The fourth root of the topological susceptibility $\chi_{top}$
as function of $T$ on the $20^3\times6$ lattice.
Left: separately for $L > 0$ and $L < 0$.
Right: comparison of our final result with
that of Alles et al.~\cite{DiGiacomo} and Edwards et al.~\cite{Edwards}.  }
\label{fig:fig2}
\end{figure}

Our first task was the search for exact zero modes. Their number is
related to the total topological charge $Q_{top}$ of the lattice configuration
through the index theorem~\cite{Laliena,Adams} :
\begin{equation}
Q_{top} \equiv Q_{index} = N_{-} - N_{+} \; ,
\label{eq:indextheorem}
\end{equation}
where $N_{-}$ and $N_{+}$ are the numbers of zero modes $\psi_{i0}$ with
negative and positive chirality,
$\psi^{\dagger}_{i0}~\gamma_5~\psi_{i0}=\pm 1$,
respectively.
Actually, we found only configurations with $N_{+}$ or $N_{-}$ or both
vanishing.
For the lattice ensembles the average topological charge
$\langle Q_{top} \rangle$ should vanish,
while the second moment $\langle Q_{top}^2 \rangle$
measures the strength of global topological fluctuations.
The topological susceptibility is measured as
\begin{equation}
\chi_{top} \equiv \frac{\langle Q_{top}^2 \rangle}{V} \; ,
\label{eq:susceptibility}
\end{equation}
where $V$ is the four-dimensional lattice volume in physical units.
In Fig.~\ref{fig:fig1} (left) we
show a histogram of the topological charge
in the confinement phase, close to the transition.
Fig.~\ref{fig:fig1} (right) shows the corresponding
histogram for a temperature higher up in the deconfinement phase.
The histogram in the confinement phase can be approximately
fitted by a Gaussian
distribution, in the deconfinement phase the Gaussian fit
is of more poor quality.
Let us now discuss the topological susceptibility as function of temperature.
In Fig.~\ref{fig:fig2} (left) we show the topological susceptibility
$\chi_{top}$ separately for the positive ($L > 0$) and the negative Polyakov
loop sector ($L < 0$) over the full temperature range.
At temperatures  $ T \le 1.31~T_c$ we have obtained both sectors in the course
of the usual Monte Carlo sampling.
Flips between the sectors happened even at $T>T_c$ due to finiteness of our
volume. For $T=1.55~T_c$ and $2.0~T_c$ we have not observed such flips
and stayed within the $L > 0$ sector. We then generated configurations
for the $L < 0$ sector in the following way. We applied a flip transformation
changing the sign of the Polyakov loop to an equilibrium configuration
with $L > 0$ and then used the Monte Carlo algorithm to produce a necessary
number of configurations for the other sector.
No flips back to the $L > 0$ sector have been observed during these runs.
Taking all temperatures into account, there is no systematic influence of
the sign of the Polyakov loop or, what is equivalent, of the type of temporal
boundary conditions, on the topological susceptibility.
This conforms to the observation in Ref.~\cite{Pullirsch},
for quenched $SU(3)$ configurations at $T=0$ selected to have topological
charge $Q=\pm 1$,
that the content of zero modes does not change with a continuous, complex-valued
factor used to implement non-trivial temporal boundary conditions in the Dirac
operator~\footnote{We stress, however, that this property is not guaranteed
for lower $\beta$ and for the Wilson action~\cite{calorons}.}.

The measurements for the two signs of the Polyakov loop
agree with each other at all $T$ within two standard deviations.
In Fig.~\ref{fig:fig2} (right) we compare our final averages for
$\chi_{top}(T)$, which include for $T<T_c$ all configurations and for $T>T_c$
only the subsample with $L > 0$,
with the results of Alles {\it et al.}~\cite{DiGiacomo}~\footnote{Of course,
our results imply that to calculate the topological susceptibility
one could average over all Polyakov loop sectors.}.
These authors have presented the values of
$10^{-4} \times \chi_{top}/\Lambda^{4}_L$
for various values of $\beta$ for Wilson's action representing different
temperatures. We took $\Lambda_{L}=14.15(42){\rm~MeV}$~\cite{DiGiacomo}
and extracted their susceptibility values $\chi_{top}(T)$ from these data.
The topological susceptibility is slowly decreasing with increasing
temperature
for both sets of data. Notice that the overlap definition of $Q$ results in a
systematically higher susceptibility than the improved field theoretic
definition employed by the Pisa group.
Additionally we show results obtained in Ref.~\cite{Edwards}.
We took the data for the topological charge from their Table IV for
temperatures $T/T_c=1.0, 1.4, 2.0$ and computed the respective
susceptibilities. Statistical errors were computed by the bootstrap method.
One can see that the results from Ref.~\cite{Edwards} agree well with ours
for $T/T_c=1$   and $T/T_c=1.4$ but slightly disagree for $T/T_c=2.0$.

\section{The spectral density and chiral symmetry restoration in
different $Z_2$ sectors}
\label{sec:spectral_density}

\begin{figure}
\centerline{\includegraphics[scale=0.35,angle=270]{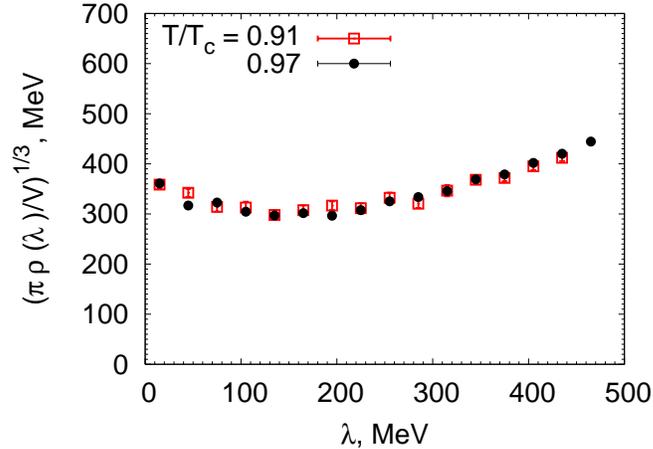}}
\caption{The third root $(\pi \rho(\lambda)/V)^{1/3}$ of the spectral
density of non-zero eigenmodes of the overlap Dirac operator (at
$\lambda=0$ approaching the third root of the quark condensate)
is shown for two temperatures $T < T_c$ as measured on the $20^3\times6$
lattice. The modes are counted with a bin size of $30 {\rm~MeV}$. }
\label{fig:fig3}
\end{figure}

The chiral condensate $\langle \bar{\psi}\psi \rangle$ is related to the
density
$\rho(\lambda)$ of the non-zero eigenvalues
$\lambda$ at $\lambda \rightarrow 0$
via the  Banks-Casher~\cite{BanksCasher} relation:
\begin{equation}
\langle \bar{\psi}\psi \rangle =- \lim_{\lambda \rightarrow 0}~\lim_{m \rightarrow 0}~\lim_{V \rightarrow \infty}~{\frac{\pi\rho(\lambda)}{V}} \; .
\label{eq:BanksCasher}
\end{equation}
The non-zero modes are globally non-chiral, but still {\it locally chiral}
over a wide part of the spectrum, and their pseudoscalar density is known 
to be correlated with lumps of the topological charge density~\cite{Ilgenfritz}.
In the chirally broken phase the required limit (\ref{eq:BanksCasher})
of $\rho(\lambda)$ is non-vanishing at $\lambda=0$~\cite{BanksCasher}.
In the chirally symmetric phase we expected vanishing $\rho(\lambda)=0$
in a finite region around the origin, in other words, that the spectrum develops
a gap. 
It cannot be excluded that the gap (an operative definition is given below)
tends to zero in the infinite-volume limit while the spectral density {\it at}
zero remains vanishing, $\rho(\lambda=0)=0$.   
For the confinement (chirally broken) phase we find indeed that
the spectral density in physical units is practically $T$ independent and
slightly increasing towards $\lambda=0$. This can be seen
in Fig.~\ref{fig:fig3} where the third root $(\pi \rho(\lambda)/V)^{1/3}$
of the spectral density is shown. At $\lambda=0$ it should approach the
third root of the quark condensate.

The eigenvalue density is defined as
\begin{equation}
\rho(\lambda)=\lim_{\Delta \lambda \to 0}
\frac{\langle N(\lambda, \Delta \lambda) \rangle}{\Delta \lambda} \, ,
\end{equation}
where $N(\lambda, \Delta\lambda)$ counts the number of eigenvalues
per configuration falling into the bin intervall
$[\lambda-\Delta\lambda/2,\lambda+\Delta\lambda/2]$. The averaging
over all configurations available is denoted by $\langle \ldots \rangle$.
In practice we have chosen a finite bin size $\Delta\lambda$
that is mentioned in each case.

We shall note that in the confinement phase,
comparing results for configurations with
average Polyakov loop $L > 0$ and $L < 0$, we found at low $\lambda$ the
quantity $(\pi \rho(\lambda)/V)^{1/3}$ to be $50 \sim 70 {\rm~MeV}$
higher for the negative Polyakov loop sector.
We observed that this difference comes mostly from configurations with
large $|L|$. This observation leads us to believe that such difference
disappears in the thermodynamic limit since it is known that $|L| \to 0$ 
in this limit.

\begin{figure}
\begin{tabular}{cc}
\hspace{-1.3cm}
  $~~~~~~~~~~~~~~L>0$ &  $~~~~~~~L<0$ \vspace{-0.5cm}\\
\hspace{-.3cm}
\includegraphics[scale=0.31,angle=270]{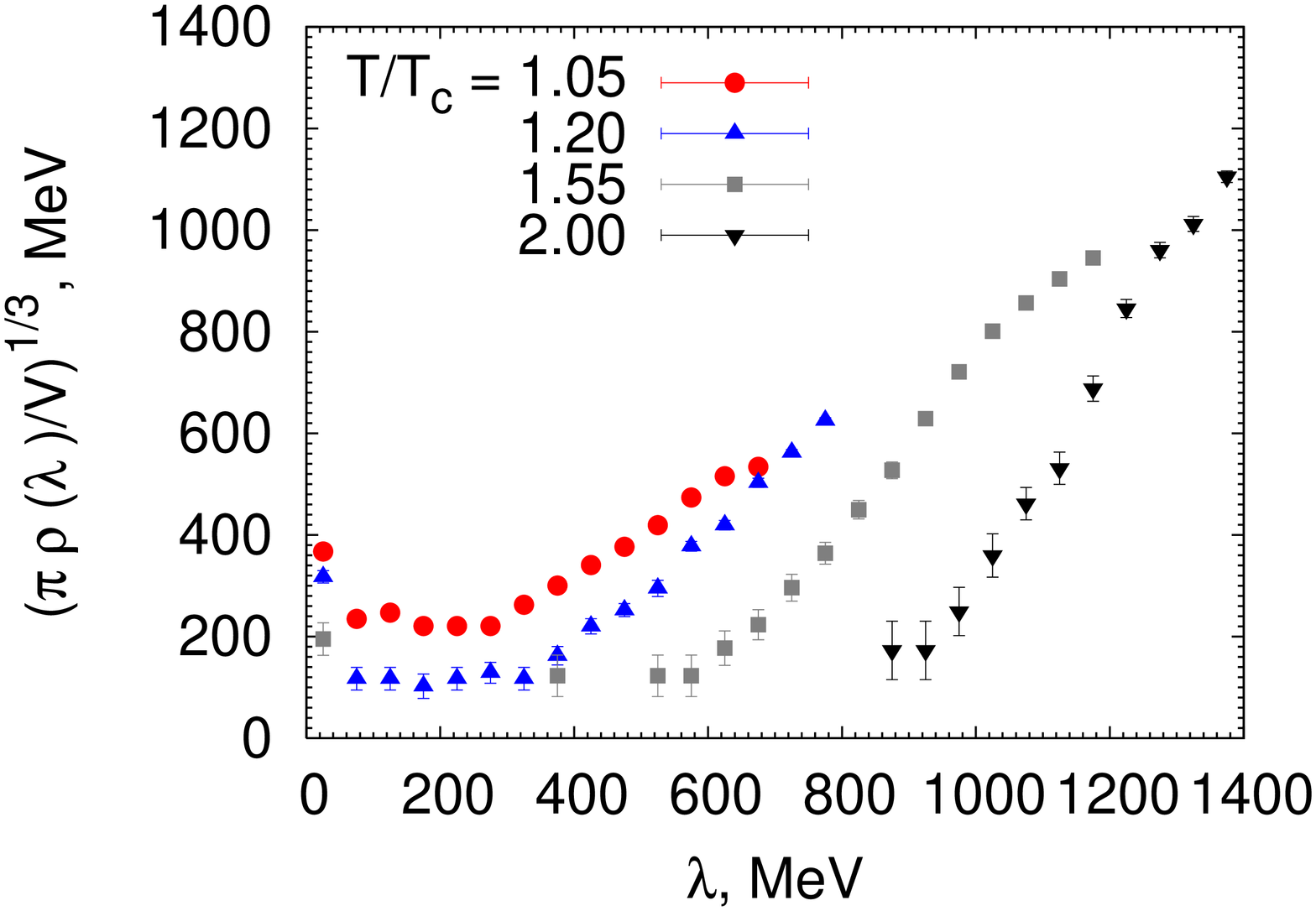} &
\includegraphics[scale=0.31,angle=270]{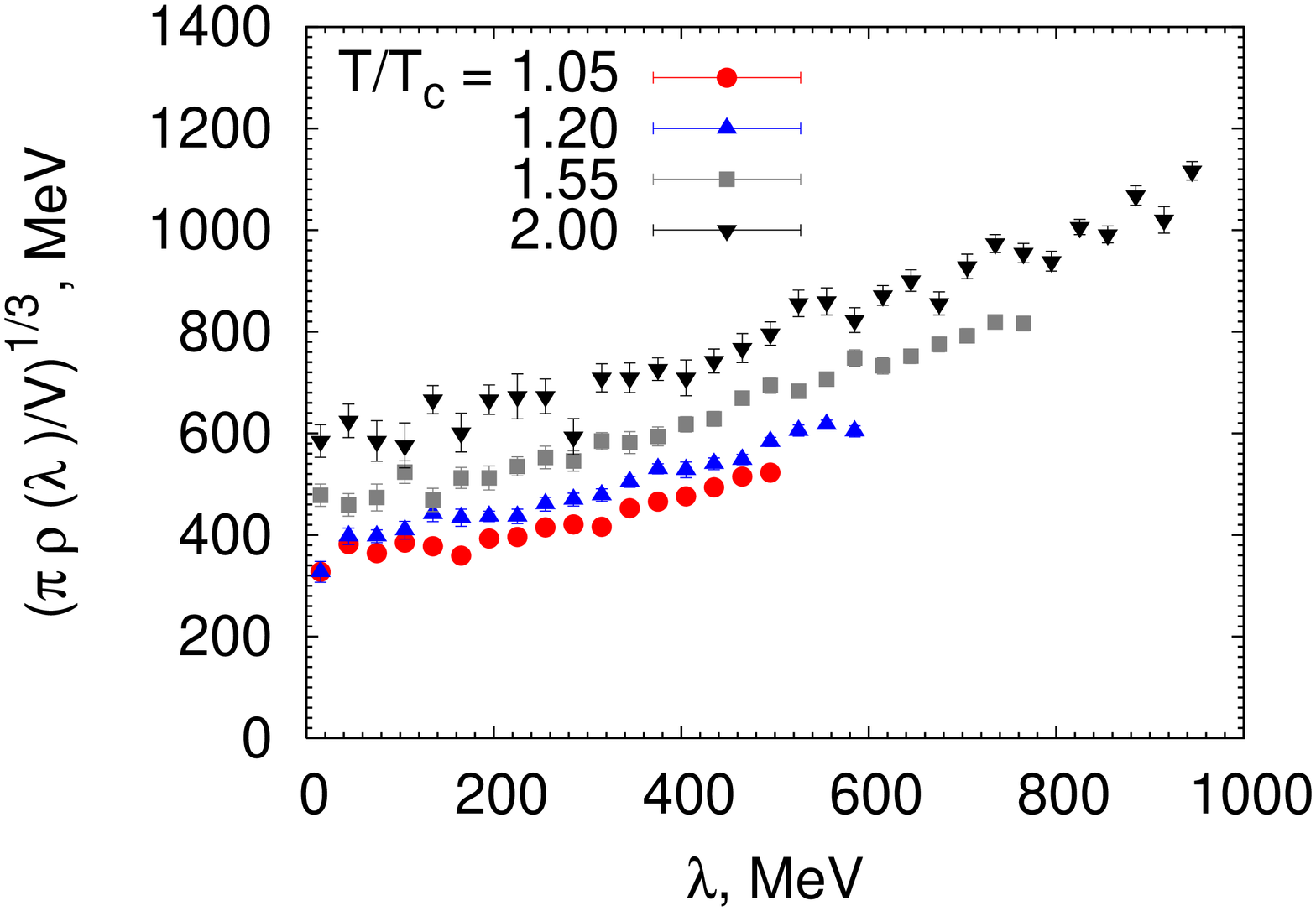} \\
\end{tabular}
\caption{The third root $(\pi \rho(\lambda)/V)^{1/3}$ of the spectral
density of non-zero eigenmodes of the overlap Dirac operator
shown for 4 temperatures $T > T_{c}$ as measured on a $20^3\times6$ lattice,
evaluated separately according to the sign of the averaged Polyakov loop:
$L>0$ (left) and $L<0$ (right).
The modes are counted for $L>0$ with a bin size of $50 {\rm~MeV}$, for
$L<0$ with a bin size of $30 {\rm~MeV}$. }
\label{fig:fig4}
\end{figure}

For the deconfinement phase, when we take only configurations
with an average Polyakov loop $L > 0$, the left panel of
Fig.~\ref{fig:fig4}
shows that $\rho(\lambda)$ non-uniformly decreases with increasing
temperature,
indicating the decrease of the quark condensate until
a gap in the spectrum opens at $T/T_c=1.55$ and gets wider at $T/T_c=2$.
This does not completely agree with our expectation above. 
The gap, when it finally opens (for the $L>0$ sector) at $T/T_c \approx 1.5$
needs a careful investigation in the infinite-volume limit.
For configurations with $L < 0$ the right panel of
Fig.~\ref{fig:fig4} shows
that the third root of $\pi \rho(\lambda)/V$ at low $\lambda$ stays
non-zero and even grows with increasing temperature.
We checked that decreasing of the bin size down to 10 MeV does not 
uncover any gaps in the spectrum for $L<0$.

For the temperatures above $T_c$ in the sector with $L > 0$,
in the interval $\lambda \lesssim 50 {\rm~MeV}$ we observe a particular
enhancement in the spectrum of non-zero modes, that monotonously decreases
with rising temperature. We will refer to these modes as ``near-zero modes''.
A similar observation has been made by Edwards {\it et al.}~\cite{Edwards} 
for quenched configurations generated with the Wilson action. The fact that 
it reappears with the improved gauge action strongly hints that it is not 
a lattice artefact. 
The ensemble average of the lowest-lying one of these non-zero eigenvalues seems 
to be a natural definition of the spectral gap (see Sect.~\ref{sec:spectral_gap}).
The presence of the spectral enhancement, however, suggests a special, more careful
analysis with respect to its origin that will be given in Sect.~\ref{sec:Edwards} 
and will result in an alternative definition of the spectral gap.

The sector with $L > 0$ plays a particular role because, as our results indicate
in agreement with model considerations of Ref.~\cite{Stephanov}, in this sector,
together with the standard antiperiodic boundary conditions for fermions,
chiral symmetry can be restored at sufficiently high temperature. We will
furthermore see on the opposite that for the sector $L < 0$, with antiperiodic 
boundary conditions kept, chiral symmetry is not restored in the sense of the
Banks-Casher relation.

However, with dynamical fermions included the sector
with $L > 0$ is the dynamically chosen sector in the high-temperature phase.
For quenched configurations, when the fermionic boundary conditions can be
changed at will from antiperiodic to periodic boundary conditions in time, 
the role of the sectors $L > 0$ and $L < 0$ is simultaneously interchanged.

\section{The spectral gap}
\label{sec:spectral_gap}

\begin{figure}
\begin{tabular}{cc}
\hspace{-1.3cm}
  $~~~~~~~~~~~~~~L>0$ &  $~~~~~~~L<0$ \vspace{-0.5cm}\\
\hspace{-.3cm}
\includegraphics[scale=0.31,angle=270]{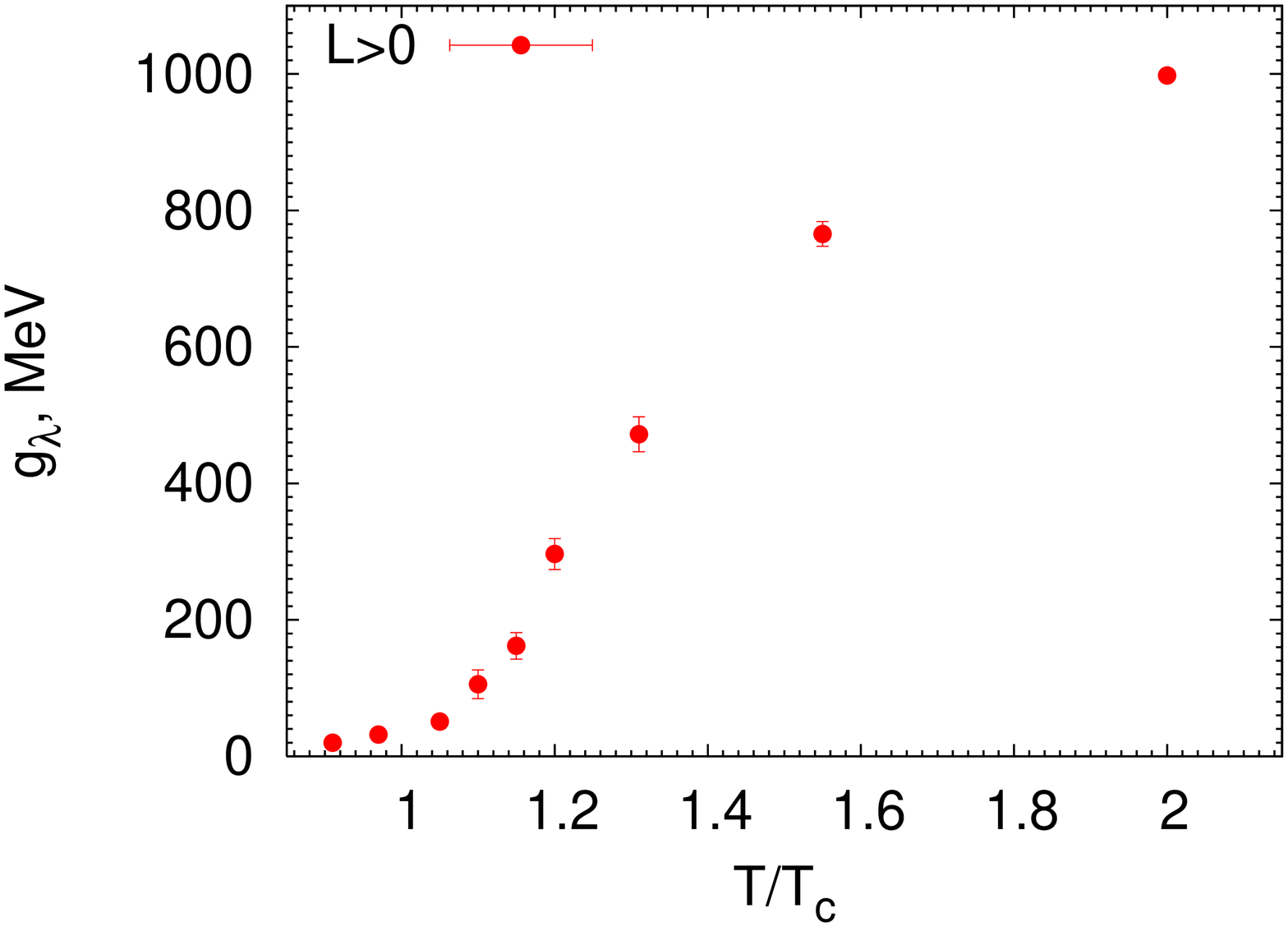} &
\includegraphics[scale=0.31,angle=270]{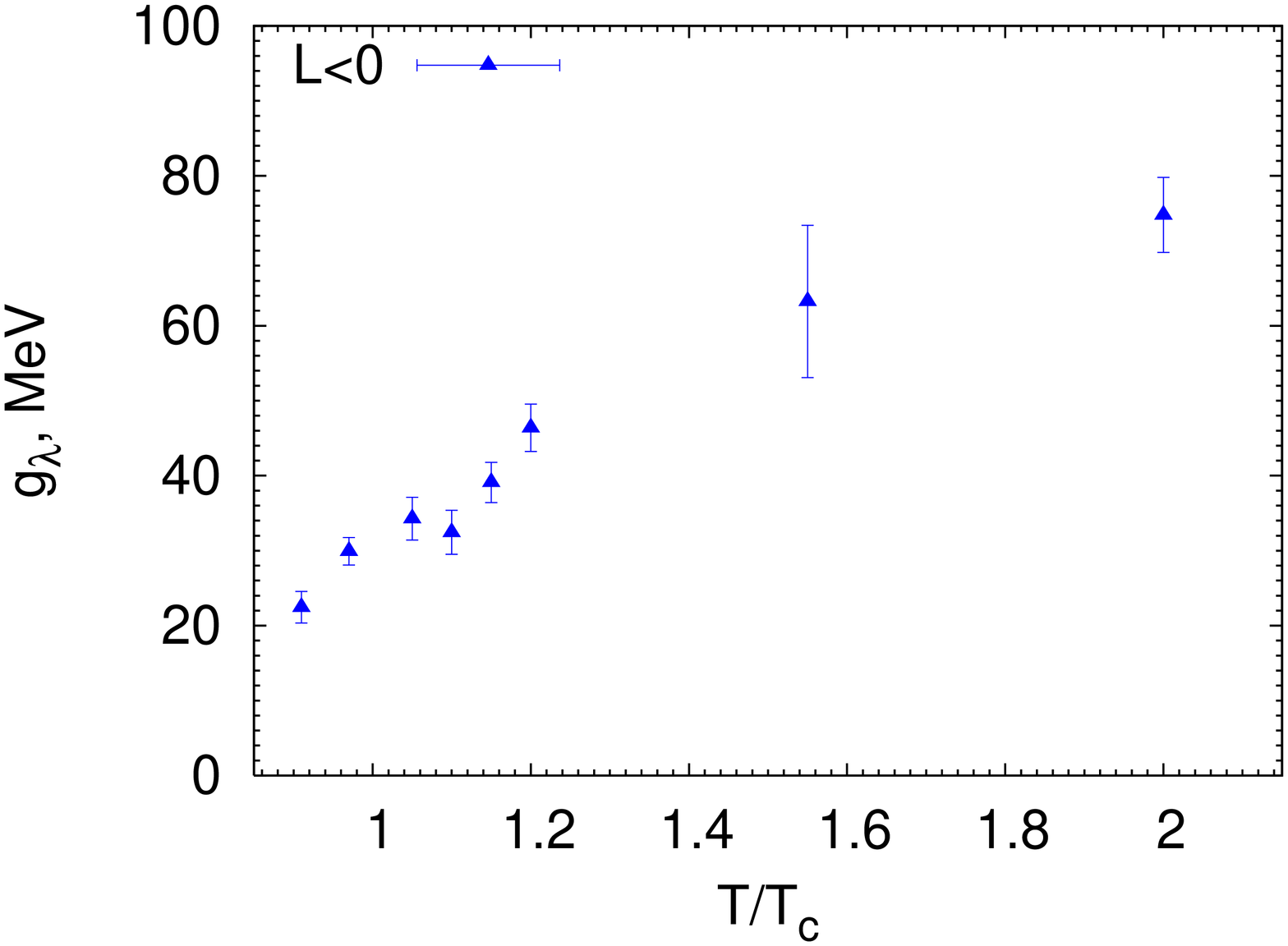} \\
\end{tabular}
\caption{The spectral gap for $SU(2)$ lattice gauge theory as function of
temperature, evaluated separately according to the sign of the averaged
Polyakov loop on a $20^3\times6$ lattice, $L > 0$ (left) and $L < 0$ (right).
Notice the scales for the gap differing by a factor 10.}
\label{fig:fig5}
\end{figure}
The spectral gap $g_{\lambda}$ has been defined as the ensemble average of the 
smallest non-zero eigenvalue. In Ref.~\cite{Gattringer} Gattringer
{\it et al.} have shown for $SU(3)$ gluodynamics that the gap, as a function
of temperature, has a different magnitude but a similar behavior for the real
and both complex sectors (defined according to the complex phase of the
averaged Polyakov loop). The phase transition -- defined by the respective
$g_{\lambda}$ becoming non-zero -- occurs at the same $T_c$.
With increasing lattice volume the gap $g_{\lambda}$ at each
temperature $T > T_c$ has the tendency to decrease, but this decrease
seemed to stop at a finite limit. The infinite-volume limit of the gap
in the three sectors has not yet been carefully analysed.

$SU(2)$ gluodynamics has only two sectors in the deconfinement phase,
distinguished by the sign of the (real-valued) averaged Polyakov loop.
We show in Fig.~\ref{fig:fig5} (left) a clearly defined and rapidly growing 
gap $g_{\lambda}$ for configurations with $L > 0$, whereas for configurations
with $L < 0$ the gap remains very small up to temperatures several times
higher than $T_c$, as can be seen from the right panel of Fig.~\ref{fig:fig5}.

Our data presented in section \ref{sec:Edwards} (see Fig.~\ref{fig:fig17}(right)) 
strongly suggest that the small gap for $L<0$  is a pure finite-volume effect 
and vanishes in the limit of spatial $V_3 \rightarrow \infty$.

Physically more important in the light of our expectations
would be to prove that the gap for $L > 0$ -- although
also decreasing with increasing volume -- has a finite limit at infinite
volume. We postpone the discussion of this question to Sect.~\ref{sec:Edwards}.
Although the separation of the near-zero modes from the definition of the gap,
that we will discuss there, will make the existence of a finite gap for all 
temperatures above $T_c$ and for all available volumes obvious, a real 
infinite-volume limit of the gap $g_{\lambda}$ has still to be explored.

\section{Localization in different parts of the spectrum}
\label{sec:localization}

\begin{figure}
\centerline{\includegraphics[scale=0.35,angle=270]{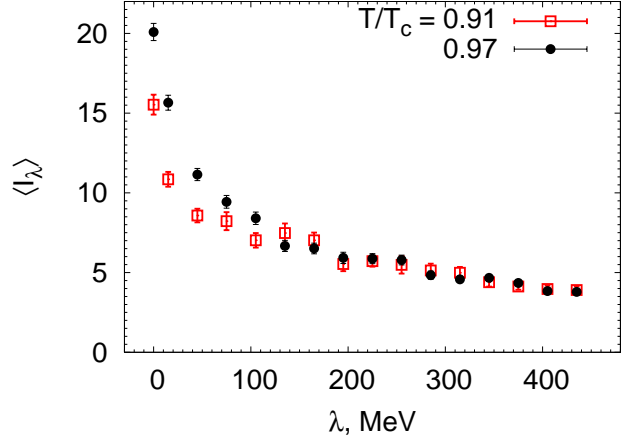}}
\caption{The IPR averaged over zero modes and over subsequent spectral bins
of width $30 {\rm~MeV}$ for two temperatures $T<T_{c}$ on the lattice
$20^3\times6$.  }
\label{fig:fig6}
\end{figure}

\begin{figure}
\begin{tabular}{cc}
\hspace{-1.3cm}
 $~~~~~~~~~~~~~~~~L>0$ &  $~~~~~~~~~L<0$ \vspace{-0.5cm} \\
\hspace{-.3cm}
\includegraphics[scale=0.31,angle=270]{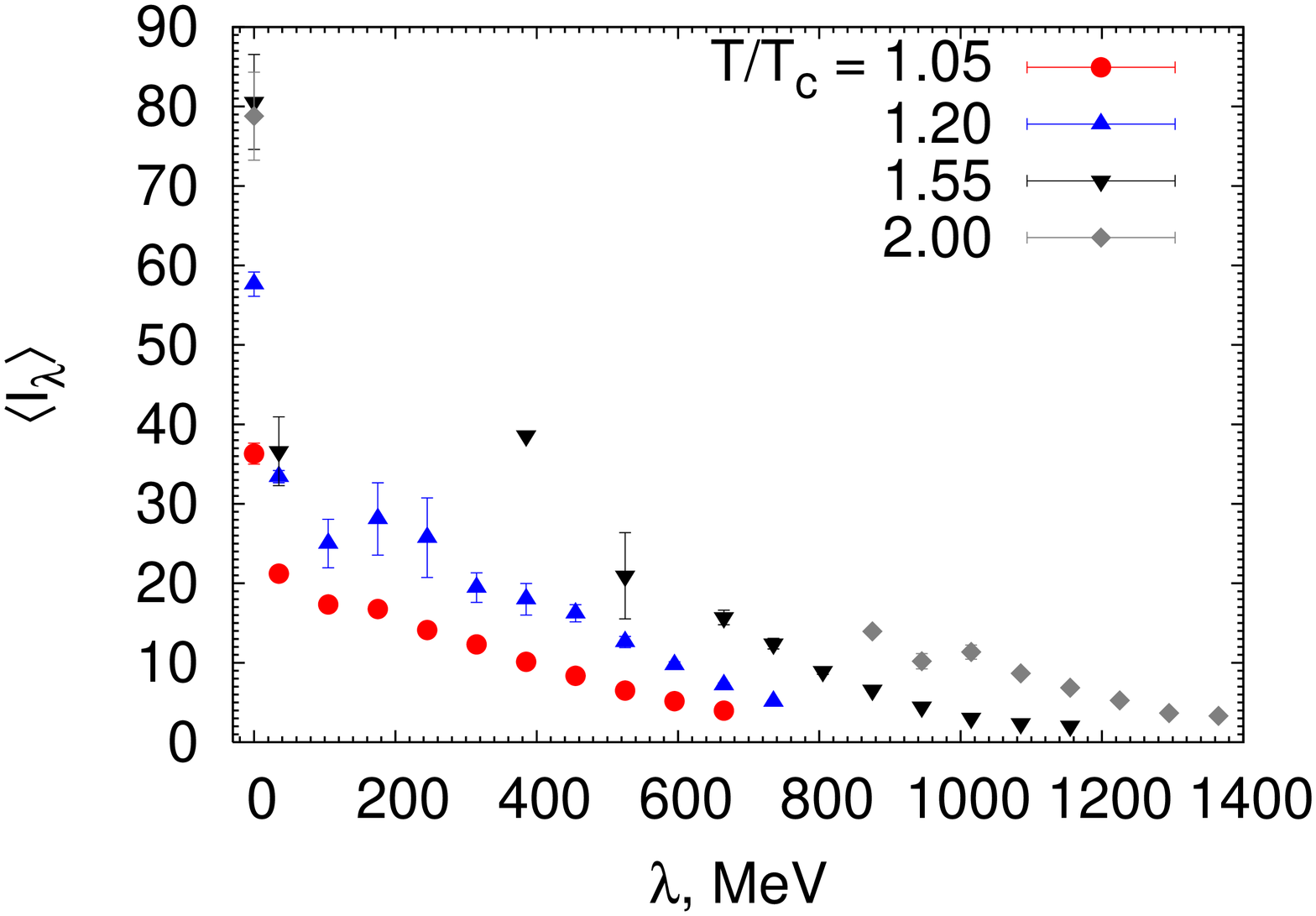} &
\includegraphics[scale=0.31,angle=270]{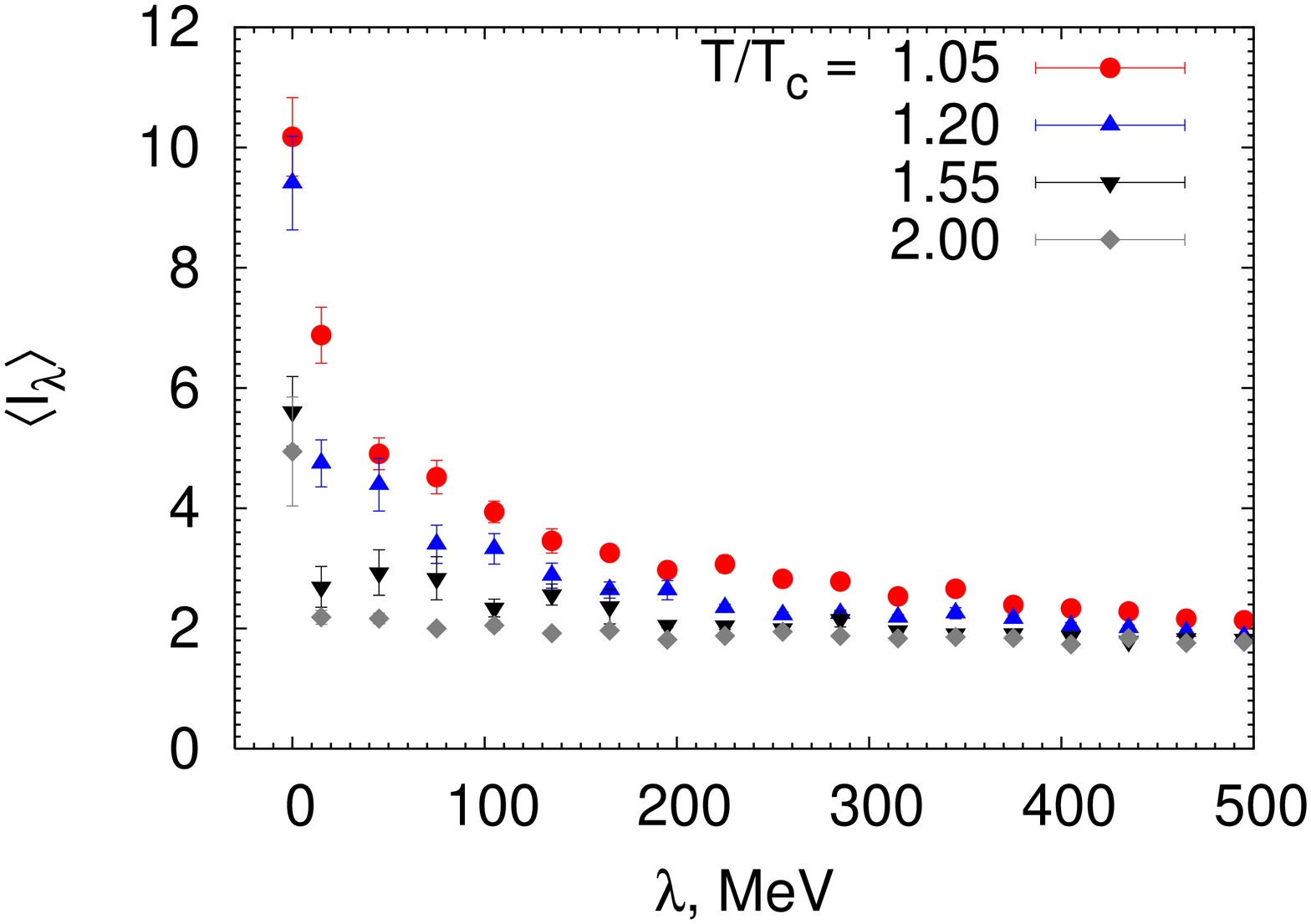} \\
\end{tabular}
\caption{The IPR averaged over zero modes and over subsequent
spectral bins of width $70 {\rm~MeV}$ (left) and $30 {\rm~MeV}$
(right), separated according to the sign of the averaged Polyakov loop
$L>0$ (left) and $L<0$ (right).	
Results are shown for four temperatures $T > T_{c}$ on the lattice
$20^3\times6$.}
\label{fig:fig7}
\end{figure}

The scalar density of an eigenmode $\psi_{\lambda}(x)$ corresponding to an
eigenvalue $\lambda$ is denoted as
$s_{\lambda}(x)=\psi_{\lambda}^{\dagger}(x)\psi_{\lambda}(x)$, such that
$\sum_{x} s_{\lambda}(x)=1$
by virtue of normalization. The inverse participation ratio (IPR)
$I_{\lambda}$ is the natural measure of the localization. For any finite
volume $V$ it is defined by
\begin{equation}
I_{\lambda}=V\sum_{x} s_{\lambda}^2(x) \; .
\end{equation}
The IPR characterizes the inverse volume fraction of sites forming
the support of
$s_{\lambda}(x)=\psi_{\lambda}^{\dagger}(x)\psi_{\lambda}(x)$.
Its value is equal to the volume $V$ for modes which are
localized on a site, scales with $V$ if it is localized on a finite
number of lattice sites and is independent of the volume if its
localization volume grows with the lattice volume.
For $T=0$ the localization of overlap eigenmodes has been first investigated
in Refs.~\cite{Gubarev,Koma,Weinberg2,Ilgenfritz}, for $T\ne0$ in
Ref.~\cite{Weinberg1,Weinberg3}. With a chirally improved Dirac operator,
the localization properties have been investigated in detail earlier by
Gattringer et al., for quenched $SU(3)$ Yang-Mills theory at $T=0$ in
Ref.~\cite{0107016} and for finite temperature in Ref.~\cite{0105023}.
The localization properties and their change with temperature are surprisingly
similar for $SU(3)$ and $SU(2)$. This applies to the overlap operator as well
as to the chirally improved Dirac operator.

In the following figures the IPR are shown at $\lambda=0$ averaged over
the zero modes, followed by the IPR averaged over the subsequent bins of
non-zero modes.
From Fig.~\ref{fig:fig6} we conclude that for the temperature near but below
$T_c$ the IPR (localization) increases with decreasing eigenvalue, more or less
in a monotonous fashion. There is no clear mobility edge.
The zero modes are even more localized exceeding the first bin by $\approx 50$ \%. 
Thus, out of the low-lying modes, the higher ones are continuously less localized.
We found that at these temperatures for configurations with large in
absolute value negative Polyakov loop, $L < 0$, the modes are less localized by a
factor $2 \sim 3$. Again, as for $\rho(\lambda)$, this difference should disappear 
in the thermodynamic limit.

In Fig.~\ref{fig:fig7} (left) we show for $L > 0$
that with increasing temperature in the deconfinement phase the average IPR
within the respective eigenvalue bins is increasing. With increasing temperature,
the effect moves to higher and higher eigenvalues $\lambda$.
This can be considered as a mobility edge moving, together with the gap,
to larger $\lambda$ in the deconfinement phase for $L > 0$.
In the negative Polyakov loop sector the IPR is
taking values at a much lower level. The temperature dependence at fixed
$\lambda$ is reverse compared to the $L > 0$ sector, and the temperature
dependence becomes stronger towards small $\lambda$. The spreading 
of the IPR values with respect to temperature is almost twice as large for 
zero modes compared to the smallest non-zero modes.

\section{Finite volume and finite lattice spacing effects
on topological susceptibility, spectral density and localization}
\label{sec:volume_and_discretization}

\begin{figure}[hpbt]
\centerline{\includegraphics[scale=0.35,angle=270]{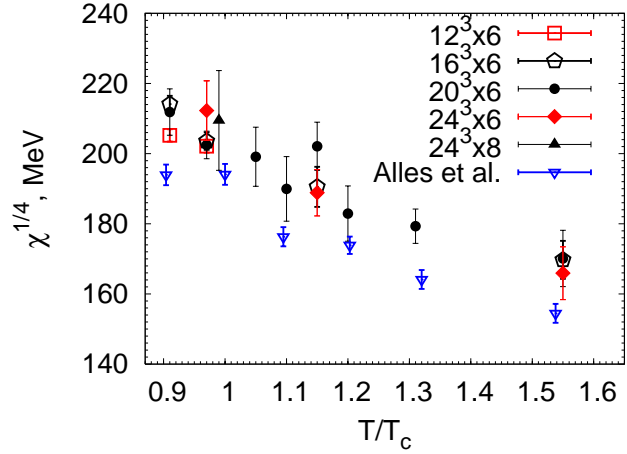}}
\caption{The fourth root of the topological susceptibility $\chi_{top}$ as
function of $T$ on $12^3\times6$, $16^3\times6$, $20^3\times6$, $24^3\times6$
and on $24^3\times8$ lattices compared with the results of Ref.~\cite{DiGiacomo}.}
\label{fig:fig8}
\end{figure}

\begin{figure}[hpbt]
\begin{tabular}{cc}
\hspace{-0.4cm}
\includegraphics[scale=0.31,angle=270]{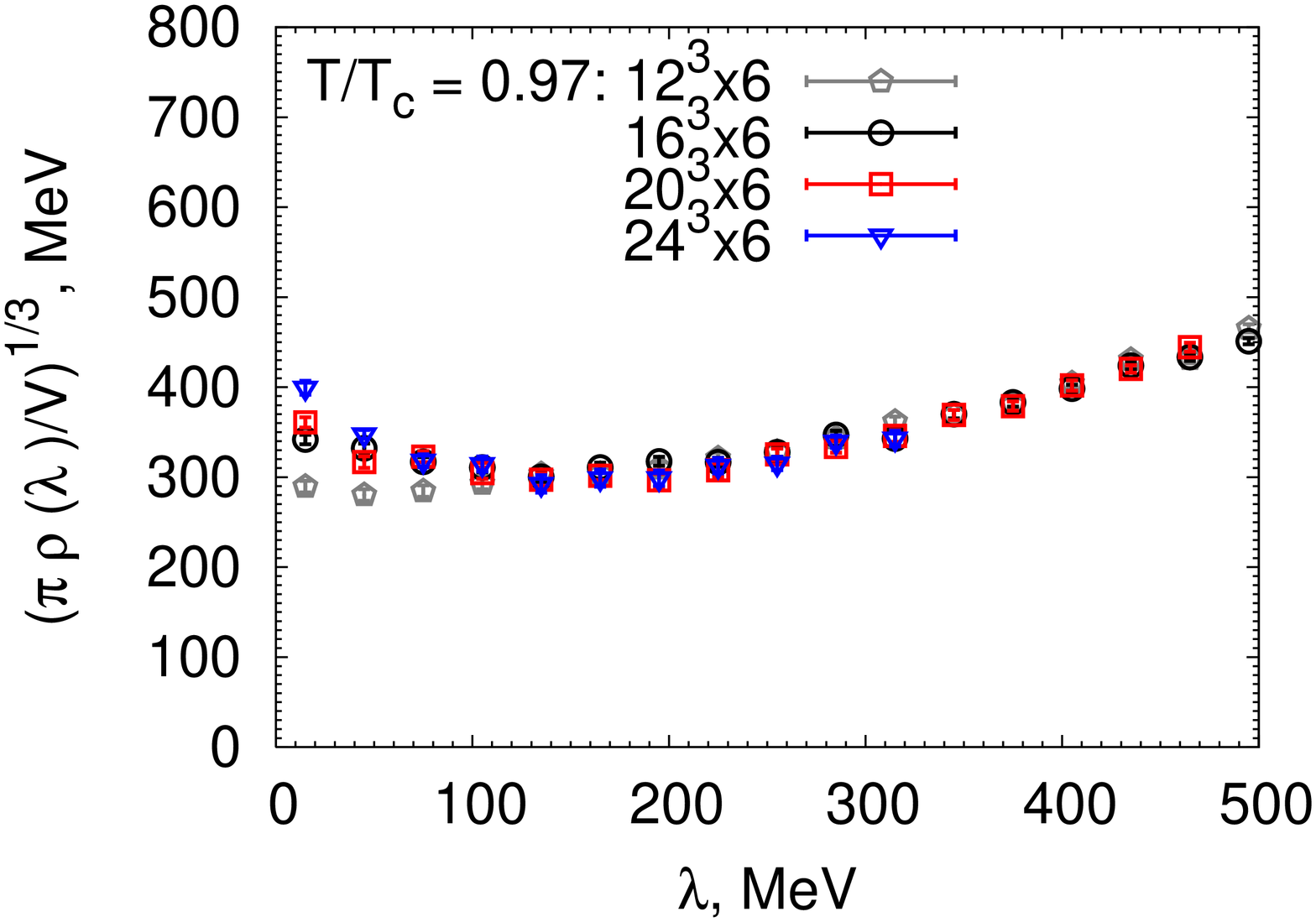} &
\includegraphics[scale=0.31,angle=270]{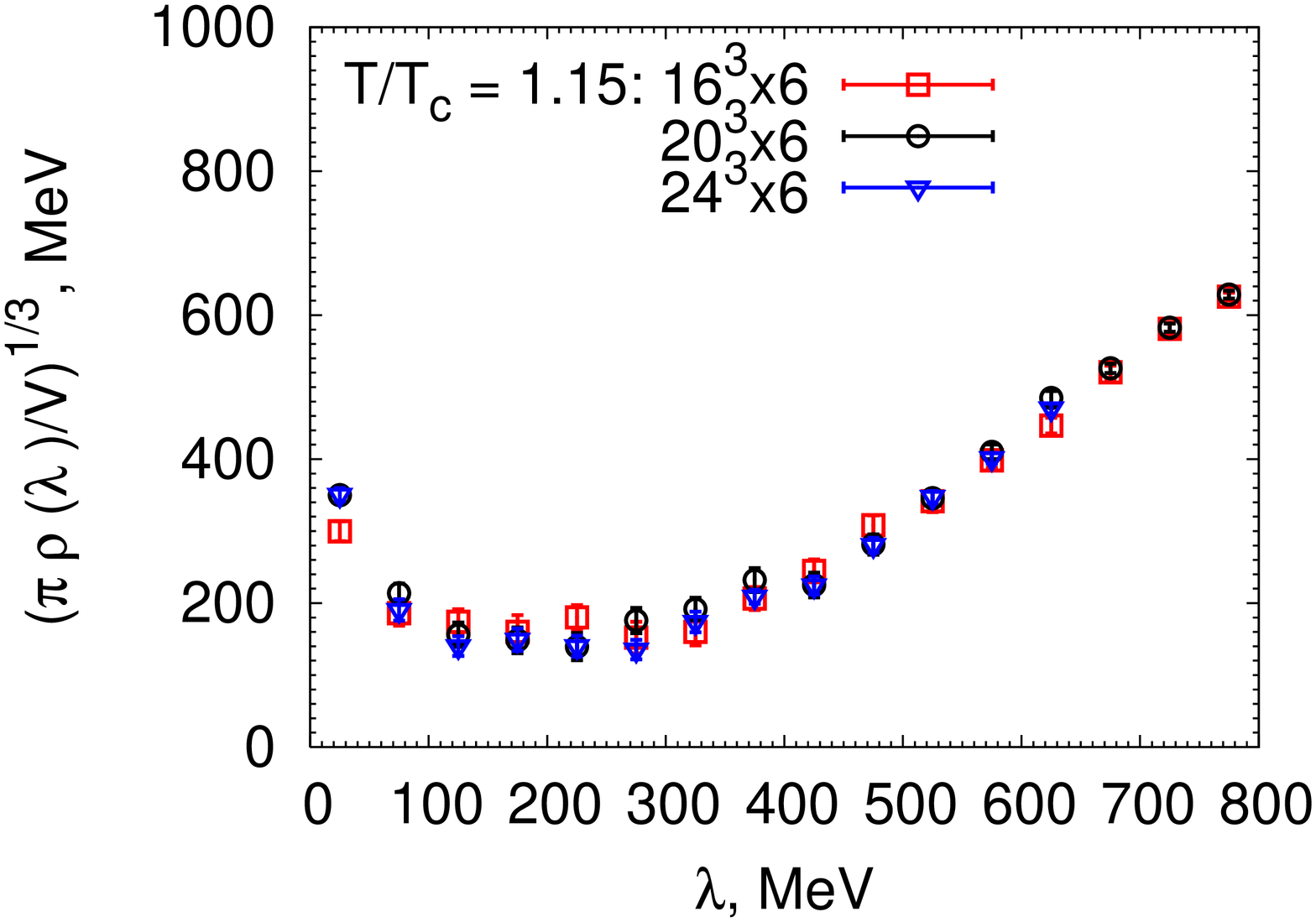} \\
\end{tabular}
\caption{The volume dependence of the third root of the averaged 
spectral density of non-zero eigenmodes of the overlap Dirac operator 
at $T=0.97~T_c$ (left) and $T=1.15~T_c$ (right). While $N_{\tau}=6$ 
in all cases, we 
compare four spatial lattice sizes with $N_s=12, 16, 20$ and $24$ 
below $T_c$ and three spatial lattice sizes with $N_s=16, 20$ and $24$ 
above $T_c$. The modes are counted with a bin size of $30 {\rm~MeV}$ 
below $T_c$ and with a bin size of $50 {\rm~MeV}$ above $T_c$. }
\label{fig:fig9}
\end{figure}

\begin{figure}[hpbt]
\centerline{\includegraphics[scale=0.35,angle=270]{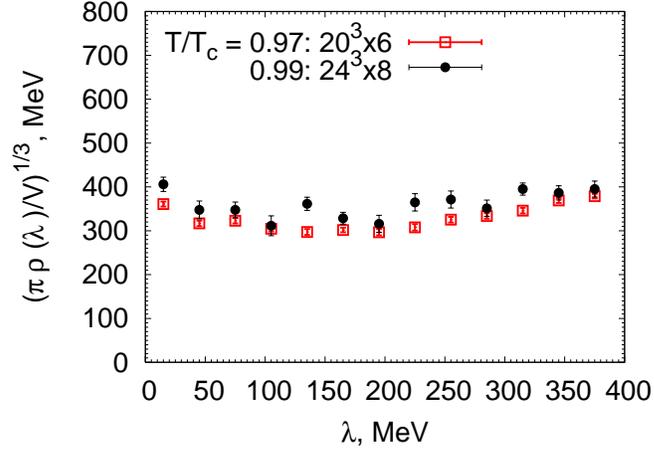} }
\caption{The third root of the spectral density of non-zero eigenmodes
of the overlap Dirac operator for two temperatures $T<T_c$ closely
approaching $T_c$ on a $20^3\times6$ and the $24^3\times8$ lattice, 
respectively. The modes are counted with a bin size of $30 {\rm~MeV}$. }
\label{fig:fig10}
\end{figure}

To study finite volume effects we have additionally generated ensembles of
configurations on $12^3\times6$,  $16^3\times6$ and  $24^3\times6$ lattices 
(see Table \ref{tbl:sim_det}).
In Fig.~\ref{fig:fig8} we compare our results for the topological
susceptibility $\chi_{top}$ obtained on all our lattices.
One can see that the finite volume effects are rather small and unimportant for this observable.
Finite lattice spacing effects seem to be
also negligible, since the $\chi_{top}$
result on $24^3\times8$ lattice is in accord with that on the
$20^3\times6$ lattice
at the temperature where we have data available (see Fig.~\ref{fig:fig8}).

From Fig.~\ref{fig:fig9} (left for $T<T_c$ and right for $T>T_c$,
restricted to the $L>0$ sector) we conclude, that the spectral density 
(if divided by the 4-volume) does not depend on the lattice volume for 
large $\lambda$, while some volume dependence is seen for 
$\lambda \lesssim 100$ MeV for $T<T_c$ (Fig.~\ref{fig:fig9} left).
Fig.~\ref{fig:fig10} demonstrates a weak, if any, 
dependence of the spectral density in the confinement phase
when the transition temperature is approached.

\begin{figure}[hpbt]
\begin{tabular}{cc}
\hspace{-.3cm}
\includegraphics[scale=0.31,angle=270]{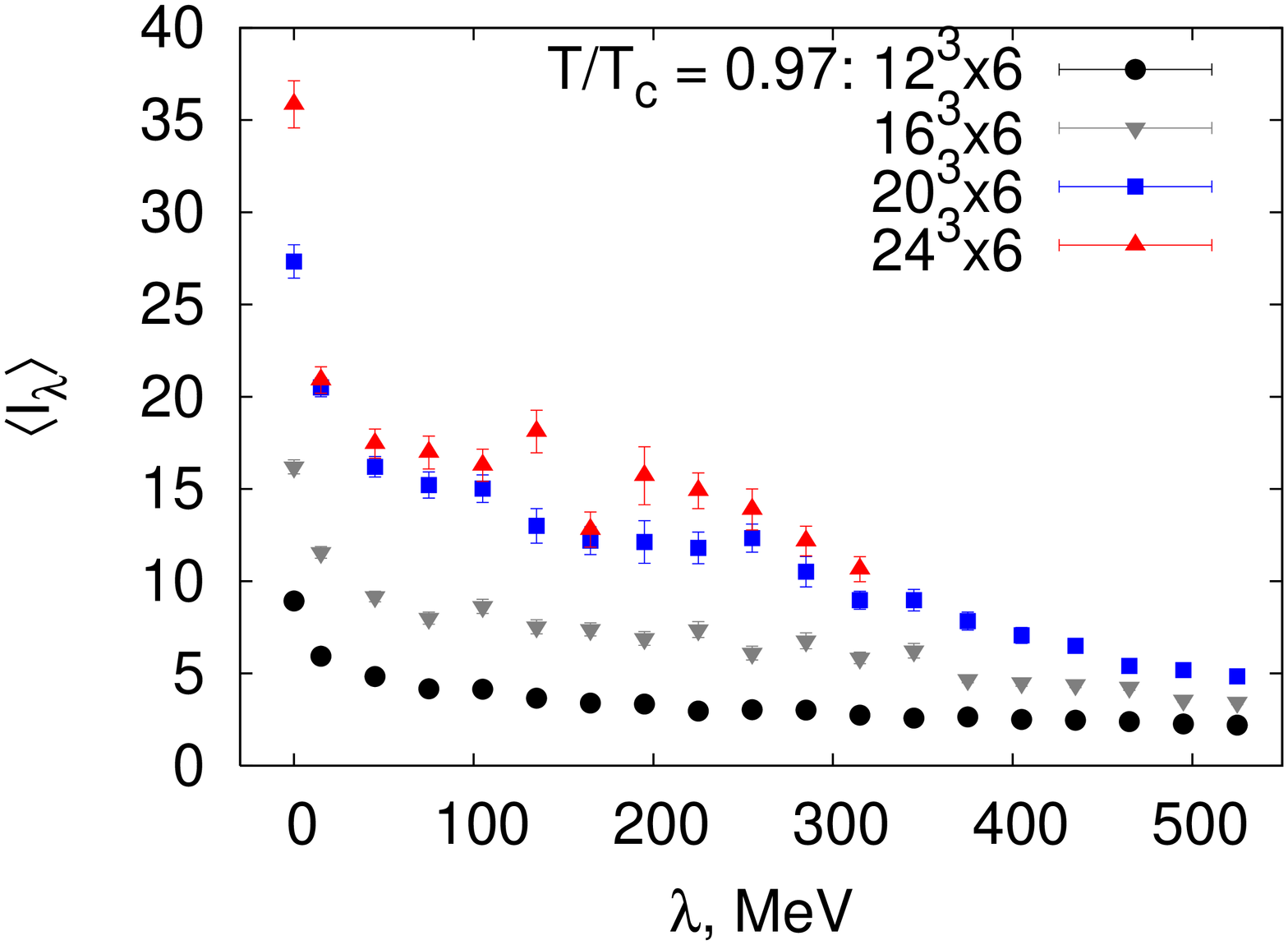} &
\includegraphics[scale=0.31,angle=270]{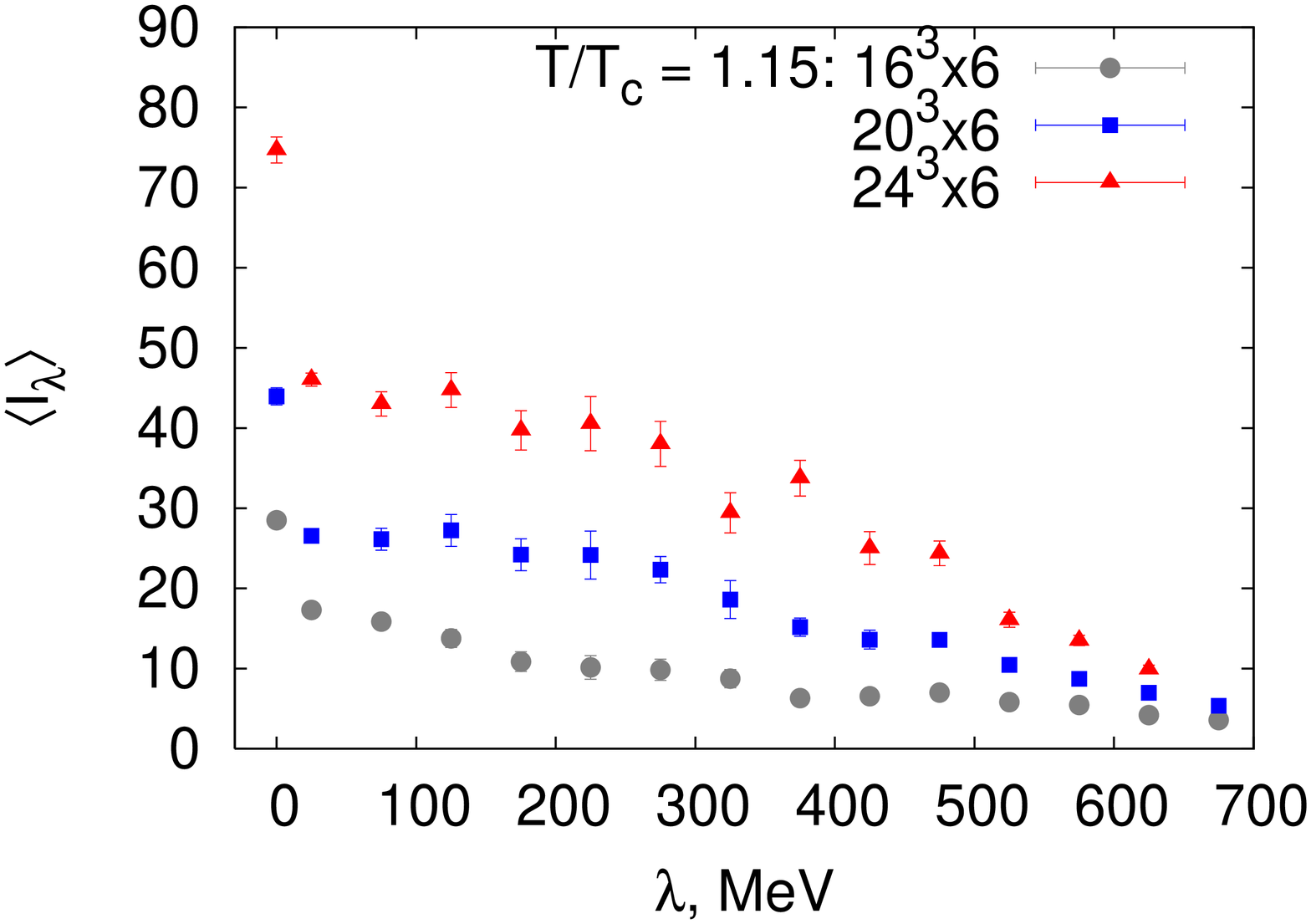} \\
\end{tabular}
\caption{The IPR averaged over zero modes and over spectral bins
of width $30 {\rm~MeV}$ (left) and $50 {\rm~MeV}$ (right).
Left: comparison for $T=0.97~T_c$ ($N_{\tau}=6$) between four spatial
volumes with $N_s=12,16, 20$ and $24$.
Right: comparison for $T=1.15~T_c$ in the $L>0$ sector between
three spatial volumes, on $16^3\times6$, $20^3\times6$ and $24^3\times6$
lattices.}
\label{fig:fig11}
\end{figure}

\begin{figure}[hpbt]
\centerline{\includegraphics[scale=0.35,angle=270]{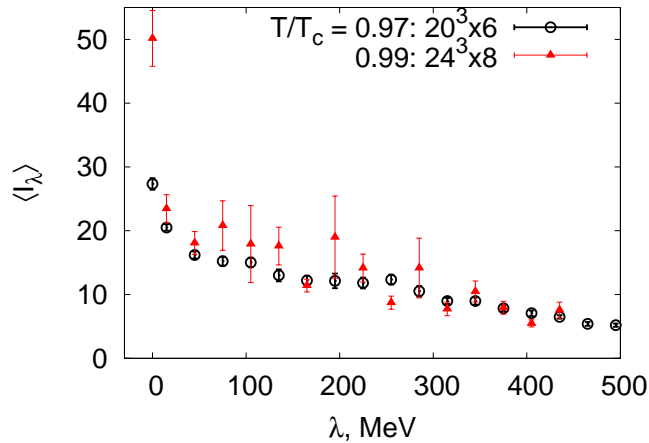}}
\caption{The IPR averaged over zero modes and over spectral bins
of width $30 {\rm~MeV}$ checked for lattice spacing effects by
comparison between the $20^3\times6$ lattice and the $24^3\times8$
lattice for two temperatures $T<T_c$ close to each other.}
\label{fig:fig12}
\end{figure}

Coming now to the localization properties of the modes, we conclude from
Fig.~\ref{fig:fig11} that the IPR with increasing spatial
volume increases (comparing lattice sizes from $12^3\times6$ to $24^3\times6$).
This applies both to the confinement (left) and the deconfined (right) phases.
In Fig.~\ref{fig:fig12} we show, comparing the $20^3\times6$
with the $24^3\times8$ lattice (at approximately the same temperature
and with a ratio of physical three-dimensional volumes of 0.86)
that the lattice spacing effects are small.
However, this is not the case for the zero modes. Their IPR nearly doubles
when the lattice spacing $a$ decreases by a factor 6/8. This corresponds
to a dimensionality of the zero modes close to $d \approx 2$~\cite{Koma}.

\section{Comparison with Edwards {\it et al.}}
\label{sec:Edwards}

As it has been said before, the overlap operator spectrum at finite temperature
has been studied both in $SU(2)$ and $SU(3)$ gluodynamics by
Edwards {\it et al.}~\cite{Edwards}.
In $SU(2)$ case they made simulations at $T/T_c=1.0$, $1.4$ and $2.0$ .
It has been found that for all temperatures considered the spectrum 
(for antiperiodic boundary conditions and positive Polyakov loop)
consisted of two parts: a group of unexpected (and unobserved with staggered 
fermions !) near-zero modes, {\it i.e.} modes with small eigenvalues below 
$0.05/a$ with a spectral density decreasing
as a function of $\lambda$, and the rest with larger eigenvalues above
$0.05/a$ with a spectral density increasing with increasing $\lambda$.
There was a large gap found between these two parts. 
The authors of Ref.~\cite{Edwards} assumed
that the small eigenvalues (the near-zero modes) are due to
instanton-antiinstanton pairs
and presented some evidence supporting this assumption. We prefer to call
them more generally ``topological objects'' in what follows.
In particular, Edwards {\it et al.} found that the number of small
(near-zero) modes $n_{sm}$, combined with the number of zero modes $n_0$,
is equal to the
number of level crossings in the spectral flow of $H_W$, pointing towards a
common origin. They stated that $\langle Q^2 \rangle$ agreed well with the
average $\langle n \rangle$ where $n = n_{sm} + n_0$. We will find a similar
remarkable agreement.
Furthermore, it was found that the distribution of $n = n_{sm} + n_0$ is
in good agreement with a Poisson distribution
\begin{equation}
 P(n;\langle n \rangle) =
 \frac{e^{-\langle n \rangle}}{n !}
 \langle n \rangle^n \; ,
\label{eq:poisson}
\end{equation}
where $\langle n \rangle$ is the ensemble average.

\begin{figure}[hpbt]
\centerline{\includegraphics[scale=0.50]{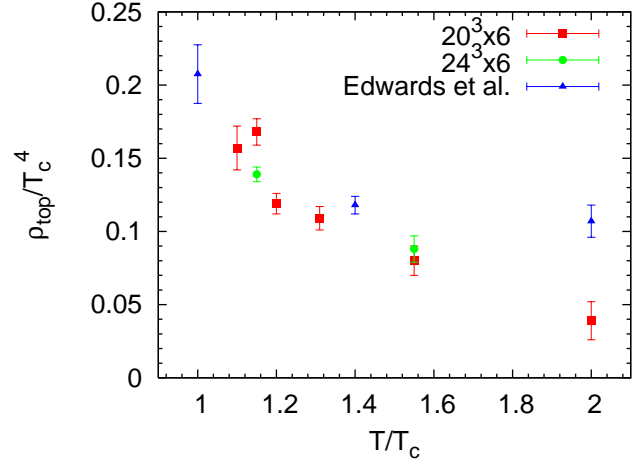}}
\caption{The density of topological objects (see text) in units of $T_c^4$ as
a function of temperature $T$, from our data obtained for two 3-volumes
at $T>T_c$ and from the data of Ref.~\cite{Edwards}.}
\label{fig:fig13}
\end{figure}
Our Fig.~\ref{fig:fig4} (left) demonstrates that we have also observed the
two parts of the eigenvalue spectrum mentioned above, although these two parts
are not separated by a gap for $T/T_c<1.55$, rather by a plateau of low
spectral density.
We therefore defined a boundary between the two parts of the spectrum
for $T/T_c < 1.55$ at 150 MeV which is approximately located at the minimum
of the distribution $\rho(\lambda)$. We checked that all our results
discussed below vary only within error bars when we vary this border value
by 50 MeV.

\begin{figure}[hpbt]
\centerline{\includegraphics[scale=0.50]{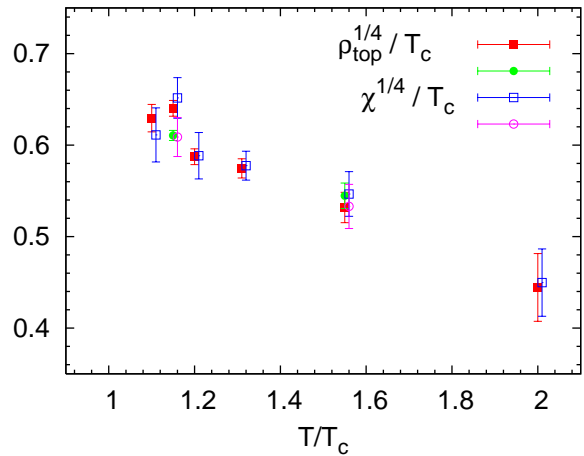}}
\caption{Comparison of $\rho_{top}^{1/4}/T_c$ (full symbols) and  $\chi^{1/4}/T_c$
(empty symbols) for $20^3\times 6$ (squares) and $24^3\times 6$ (circles) lattices.}
\label{fig:fig14}
\end{figure}

Following the discussion in Ref.~\cite{Edwards} we studied
properties of $\langle n \rangle$ and of the probability distribution
$P(n;\langle n \rangle)$ defined above.
In Fig.~\ref{fig:fig13}
we present the density $\rho_{top}$ defined as
\begin{equation}
 \rho_{top} = \frac{\langle n \rangle}{V}
\end{equation}
as a function of the temperature. For comparison, we show data from
Ref.~\cite{Edwards} which we took from their Table II.
We computed the respective statistical errors by applying the bootstrap method
to data collected there.
We present in Fig.~\ref{fig:fig13} our results for the
$24^3 \times 6$ lattice as well.
There is good agreement between our results and results of
Ref.~\cite{Edwards}
for smaller temperatures while data clearly disagree for the largest
temperature $T/T_c=2.0$~\footnote{For this temperature there is a discrepance
with the results of Ref.~\cite{Edwards} also in the case of the topological
susceptibility.}. One can see no essential volume dependence for this quantity
which has been concluded by Edwards {\it et al.} as well.
It is also seen that the density $\rho_{top}$ is slowly decreasing with
increasing temperature. The decrease with temperature is similar to that
of the topological susceptibility.
Numerically we indeed found nice agreement between $\chi$ and $\rho_{top}$
as can be seen from Fig.~\ref{fig:fig14}.

\begin{figure}[hpbt]
\centerline{\includegraphics[scale=0.50]{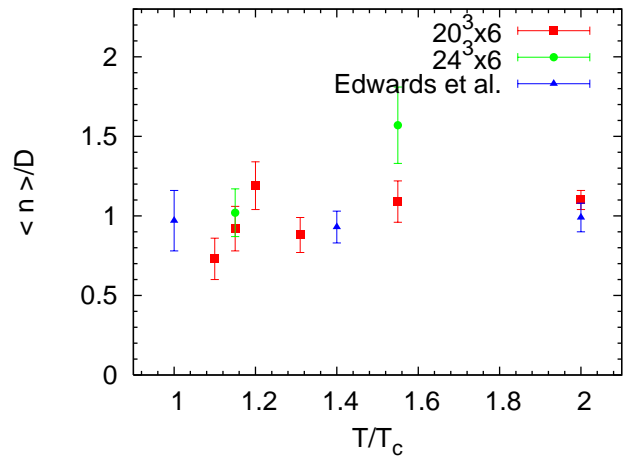}}
\caption{The ratio of the average number of all topological objects
$\langle n \rangle$ to its variance as a function of temperature $T$,
from our data obtained for two 3-volumes at $T>T_c$ and from the data
of Ref.~\cite{Edwards}.}
\label{fig:fig15}
\end{figure}

In Fig.~\ref{fig:fig15} the ratio $\langle n \rangle/D$ is depicted,
where $D$ is the variance of the multiplicity $n$ of topological objects
including near-zero and
zero modes. For the Poisson distribution exactly $\langle n \rangle = D$ holds.
One can see that our results agree with this equality except for one data point.
Thus we confirm the earlier observation of Ref.~\cite{Edwards} that the
distribution of $n$ is consistent with the assumption of mutual independence
between the respective topological objects.

\begin{figure}[hpbt]
\centerline{\includegraphics[scale=0.50]{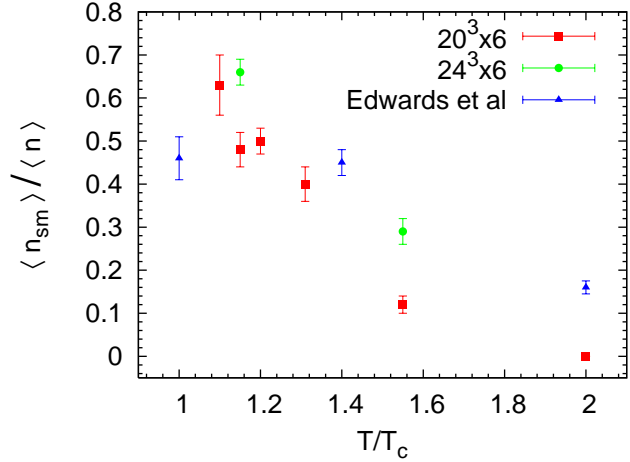}}
\caption{The ratio of the average number of small non-zero modes
$\langle n_{sm} \rangle$ to the average number of all topological objects
$\langle n \rangle$ as a function of temperature $T$, from our data
obtained for two 3-volumes at $T>T_c$ and from the data of
Ref.~\cite{Edwards}.}
\label{fig:fig16}
\end{figure}

In Fig.~\ref{fig:fig16} we show the ratio
$\langle n_{sm} \rangle / \langle n \rangle$ which indicates how the
contribution of the near-zero (small eigenvalue) modes to all topological
objects changes with volume and eventually with temperature. 
At two temperatures, $T/T_c=1.15$ and $T/T_c=1.55$, where we are in the 
position to compare two three-dimensional volumes ($N_s=20$ and $N_s=24$ 
with the same $N_{\tau}=6$) we could find this ratio rising with the volume. 
In other words, the larger the three-dimensional volume, the more near-zero
modes exist compared to the zero modes. This behavior is consistent with a 
binomial distribution proposed in Ref.~\cite{Edwards} for topological 
objects with positive and negative charge, with $n_0=|n_{+}-n_{-}|$ zero 
modes and $n_{sm}=2~min\left(n{+},n{-}\right)$ near-zero modes.
On the other hand, the share of near-zero modes among all topological 
objects seems to decrease with increasing temperature. One should have
in mind that the physical spatial volume in our setting rapidly decreases
(like $\left(N_s/(N_{\tau}~T)\right)^3$) with increasing temperature. 
For $T/T_c > 1.2$ the ratio is already less than one half, indicating 
that the volumes are already so small that the total multiplicity is 
suppressed such that there are more zero modes than near-zero modes. 

Now let us look at the problem of the spectral gap taking into account the
existence of a separate part of near-zero eigenvalues in the spectrum as
discussed above.
Alternatively to the definition of the spectral gap used in section
\ref{sec:spectral_gap} one can consider a definition $\bar{g}_\lambda$
which takes into account only the bulk non-zero modes in the spectrum, 
thus ignoring the small non-zero eigenvalues (near-zero modes) altogether. 
This alternative definition can be applied to our data at
$T/T_c=1.55$ , a temperature at which we observe a clear gap between
near-zero and larger eigenvalues.
\begin{figure}
\begin{tabular}{cc}
\hspace{-1.3cm}
 $~~~~~~~~~~~~~~~~L>0$ &  $~~~~~~~~~L<0$ \vspace{-0.5cm} \\
\hspace{-.3cm}
\includegraphics[scale=0.31,angle=270]{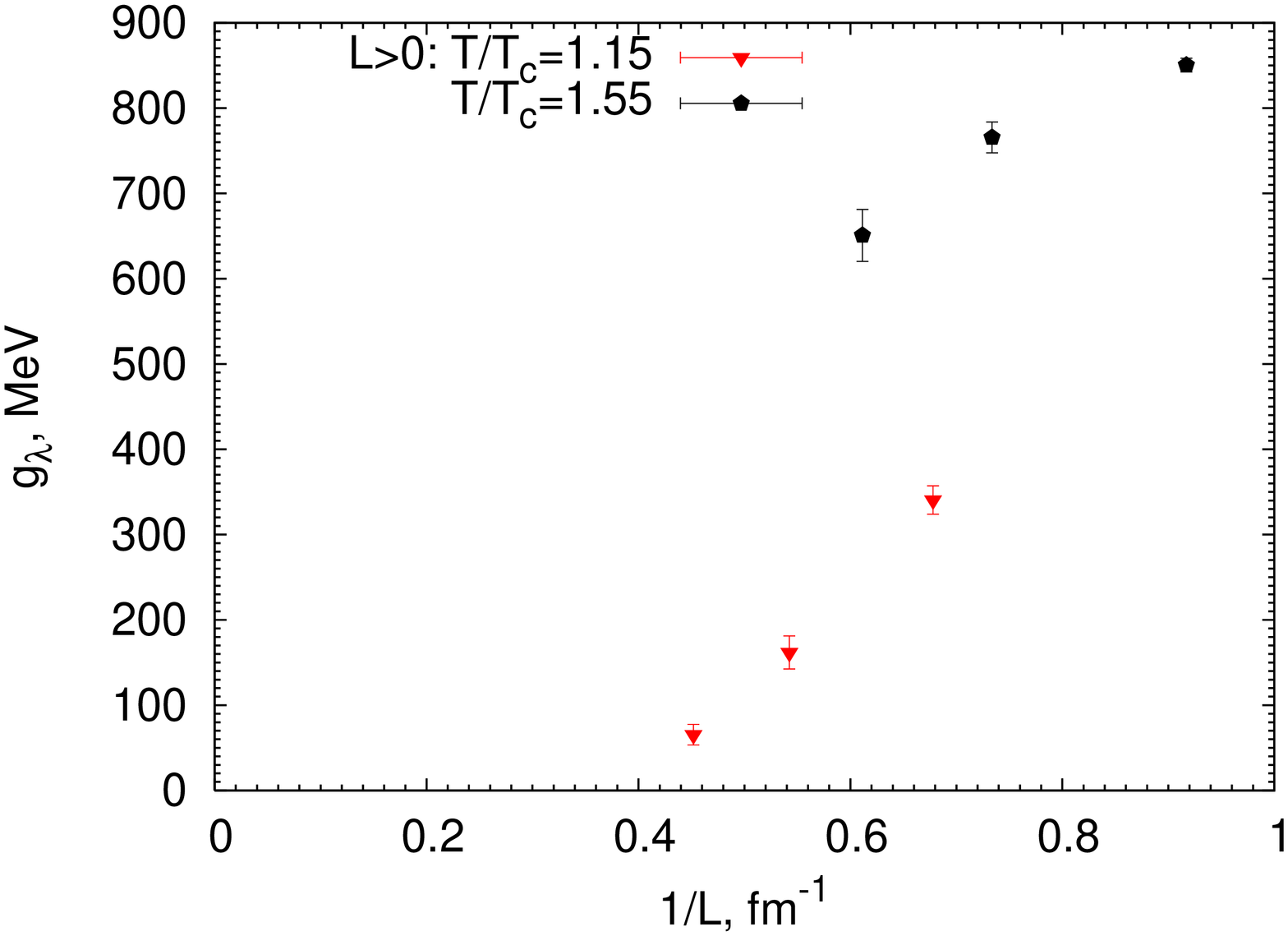} &
\includegraphics[scale=0.31,angle=270]{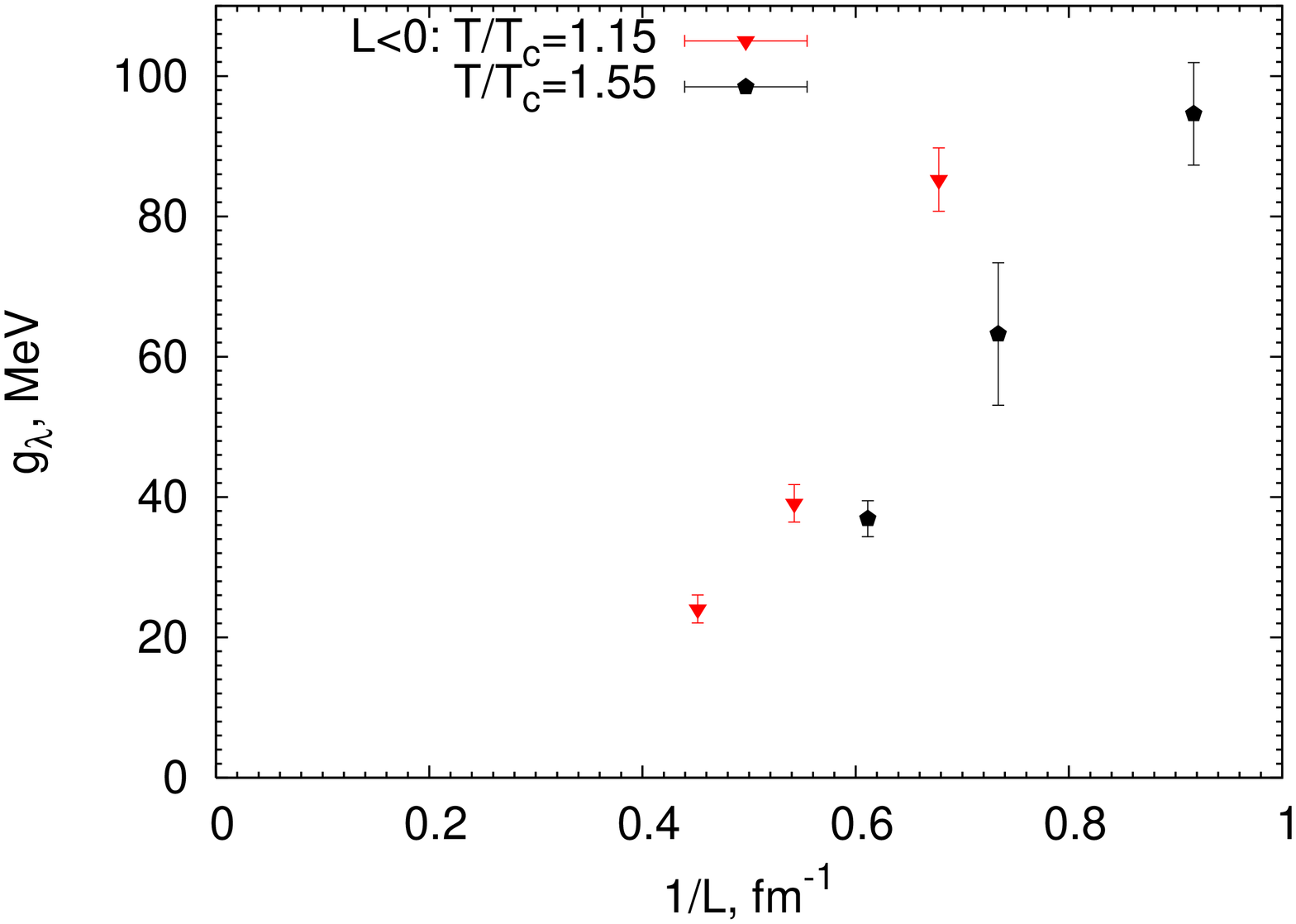} \\
\end{tabular}
\caption{The spectral gap $g_\lambda$ as defined in 
section \ref{sec:spectral_gap}
vs. inverse lattice size for $T/T_c=1.15$ and $T/T_c=1.55$,
separated according to the sign of the averaged Polyakov loop
$L>0$ (left) and $L<0$ (right).}
\label{fig:fig17}
\end{figure}
In the left panel of Fig.~\ref{fig:fig17} we show the dependence of the
conventional spectral gap, as defined in section \ref{sec:spectral_gap},
on the inverse lattice size for two temperatures in the case of $L > 0$.
The general tendency of a spectral gap is decreasing with increasing
volume. The interesting question is whether it has an infinite-volume limit
different from zero. One can see that for $T/T_c=1.15$ the spectral gap
decreases very steeply with increasing volume, starting from values like
$g_{\lambda} \approx 300 ... 400 {\rm MeV}$ on lattices of size
$L \sim 1.4 {\rm fm}$. The gap is already less than $100 {\rm~MeV}$ for
$L > 2.2 {\rm~fm}$, eventually turning to zero in the infinite-volume
limit. At higher temperature, $T/T_c=1.55$, we also see that the ``old''
gap $g_\lambda$ decreases with increasing volume, although not so fast.
Still, we cannot exclude that the fate of the spectral gap is the same
at the higher temperature.

Thus, the old definition of the spectral gap gives some support to the
conclusion made in Ref.~\cite{Edwards} that in the deconfinement
phase of $SU(2)$ gluodynamics the chiral condensate is nonzero, at least
up to some specific temperature, even in the $L > 0$ sector.

In the right panel of Fig.~\ref{fig:fig17} the conventional spectral gap
$g_\lambda$ is also shown for $L < 0$. 
Compared to the case of $L > 0$, the spectral
gap for $L < 0$ is much smaller. There is a clear tendency for $g_{\lambda}$
at $L < 0$ steeply to decrease with increasing volume. 
Data in Fig.~\ref{fig:fig17}(right) suggest that the dependence of $g_{\lambda}$
on $1/L$ is $1/L^{\gamma}$ with $\gamma>1$. Our rough estimation results
in $\gamma \approx 3$.
To make a more precise statement we need new data points for bigger
volumes. The same is valid for the data set at $T/T_c=1.15$ for $L>0$ shown
in the right panel of Fig.~\ref{fig:fig17}.
 
For the same physical 3-volume, the gap increases with increasing temperature
only in the case of $L > 0$. For the negative Polyakov loop sector the
tendency is reverse, corresponding to the increasing spectral density.

\begin{figure}[hpbt]
\hspace{-.3cm}
\includegraphics[scale=0.64,angle=0]{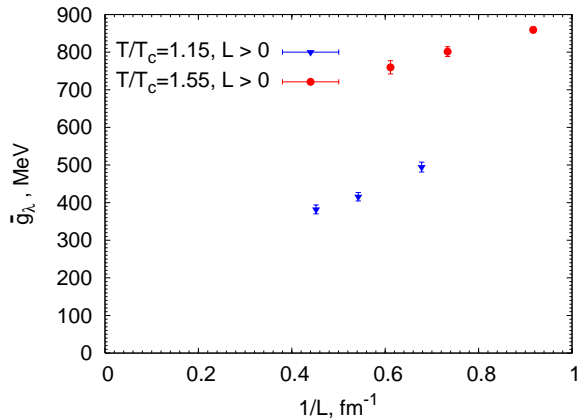}
\caption{The ``new'' spectral gap $\bar{g}_\lambda$ 
for $L>0$, as defined in section \ref{sec:Edwards}
with exclusion of the near-zero modes, is shown
vs. inverse lattice size for  $T/T_c=1.15$ and $T/T_c=1.55$.}
\label{fig:fig18}
\end{figure}
Let us now turn to Fig.~\ref{fig:fig18} where results for the ``new''
definition $\bar{g}_\lambda$ of the spectral gap, 
excluding the small eigenvalues of the
near-zero modes, are presented.
Now we see a strong indication for a convergence of the spectral gap 
$\bar{g}_\lambda$ to
a finite intercept at $1/L = 0$ for both temperatures. Note that
for $T/T_c=1.15$ only  eigenvalues $\lambda < 120 {\rm~MeV}$ have been excluded
in this way, a value substantially below the value of the new emerging gap.
Thus, excluding the near-zero modes leads to a definition of the spectral
gap $\bar{g}_\lambda$, 
such that it stays non-zero in the limit $V \to \infty$ for $L > 0$ at
both temperatures $T > T_c$.

Summarizing our findings presented in this section we confirm the existence,
at $T>T_c$, of particular modes with small non-zero eigenvalues which are most
probably related to topological objects, similar to the zero modes.
So it is natural to consider them separately from the other (bulk) non-zero modes.
We are planning to enter a further study of their properties in the near future.

\section{Comparison with T=0}
\label{sec:zero_temperature}

\begin{figure}[hpbt]
\centerline{\includegraphics[scale=0.35,angle=270]{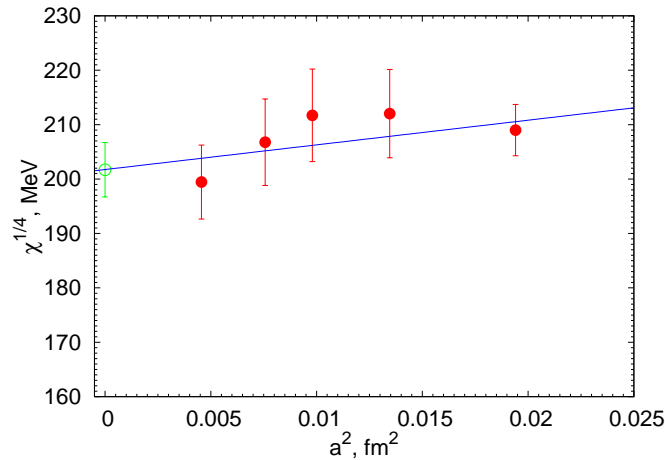}}
\caption{The fourth root of the topological susceptibility at $T=0$ as function
of the lattice spacing $a$. The empty circle shows the result of the linear
extrapolation to $a=0$.}
\label{fig:fig19}
\end{figure}

The details of simulations at zero temperature are presented
in Table~\ref{tbl:sim_det}.
The lattice coupling and the lattice size were chosen in such a way
that the physical box size was kept approximately fixed and
equal to $1.4~\mathrm{fm}$.
In Fig.~\ref{fig:fig19} we show results for the topological susceptibility
as a function of the lattice spacing squared.
The linear fit results in $\chi^{1/4} = 202(5)~\mathrm{MeV}$.
This number is in good agreement
with earlier results obtained with various
gluonic definitions of the topological
charge on lattices generated with the Wilson action. It also agrees well with
our results for $T<T_c$.  Comparing Fig.~\ref{fig:fig19} with the results
for the Wilson action~\cite{Gubarev}, obtained using the same method to compute
the topological charge as applied in the present paper,
we see that the scaling properties are better for the tadpole-improved Symanzik
action.

\begin{figure}[hpbt]
\centerline{\includegraphics[scale=0.35,angle=270]{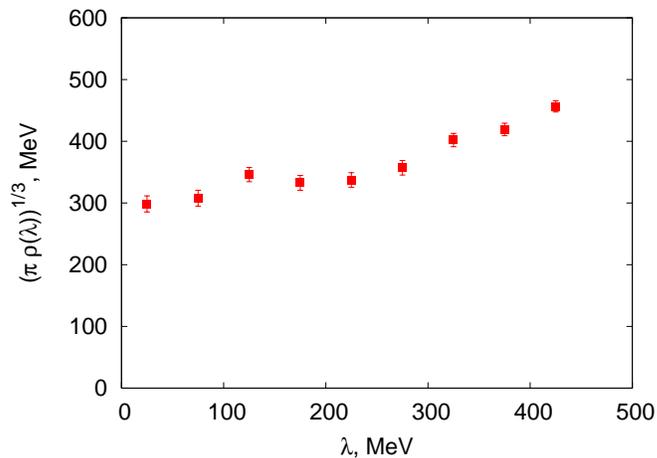}}
\caption{The spectral density obtained on the $14^4$ lattice with $\beta=3.281$.
The modes are counted with a bin size of $50 {\rm~MeV}$. }
\label{fig:fig20}
\end{figure}

In Fig.~\ref{fig:fig20} the spectral density as obtained on the $14^4$ lattice
for $\beta_{\rm imp}=3.281$
is shown. This coupling value is close to one value ($\beta_{\rm imp}=3.275$)
used in the finite temperature simulations at $T \sim T_c$. One can see
that the spectral density in Fig.~\ref{fig:fig20}
is in good agreement with that shown in Fig.~\ref{fig:fig3} except for
the lowest interval $\lambda < 100~\mathrm{MeV}$.
For $T=0$, we do not see in this range of $\lambda$
the increase of $\rho(\lambda)$ with $\lambda \to 0$
that was observed approaching $T \to T_c$ from below.
Comparing with the respective result for the Wilson action~\cite{Gubarev}
we see good agreement.

\begin{figure}[hpbt]
\hspace{1cm}
\centerline{\includegraphics[scale=0.35,angle=270]{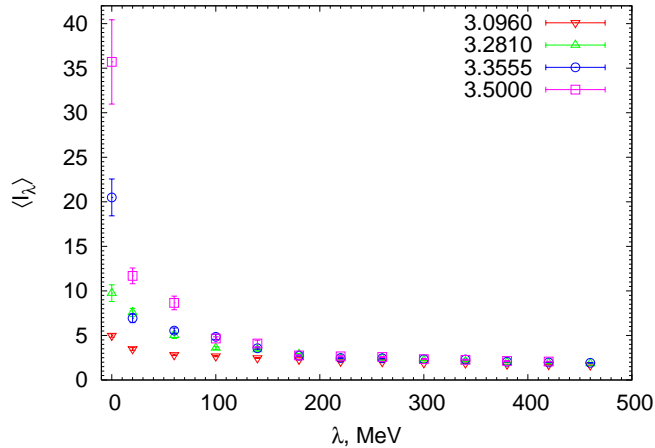}}
\vspace{0cm}
\caption{The IPR averaged over zero modes and over subsequent spectral bins
of width $30 {\rm~MeV}$ at $T=0$ for four different $\beta$-values .}
\label{fig:fig21}
\end{figure}

We  have also measured the IPR for zero temperature.
The results are presented in Fig.~\ref{fig:fig21}.
The results are again in full agreement with those of Ref.~\cite{Gubarev}.
Comparing the $T=0$ results with our observations at $T<T_c$
(see Fig.~\ref{fig:fig6}) at approximately equal $\beta_{\rm imp}$,
we see an essential difference: at non-zero temperature the IPR
close to $\lambda=0$  is substantially higher.

\section{Summary}
\label{sec:summary}

After a first overlap study of topological aspects of $SU(2)$ pure
gauge theory at finite temperature~\cite{Edwards}, based on the
Wilson gauge field action, we returned to these topics.
We have performed measurements of the topological susceptibility, 
the spectral density and the localization properties
of fermionic eigenmodes with the help of the overlap Dirac operator
in finite temperature $SU(2)$ gluodynamics, simulated with the
tadpole-improved Symanzik action. We found that
the topological susceptibility in the confinement phase
is almost independent of temperature. It starts slowly to decrease
in the deconfinement phase,
in agreement with previous results~\cite{DiGiacomo} which were
obtained with the Wilson gauge field action and an improved gluonic
definition of the topological charge density.
In contrast to the other quantities, we did not find systematic effects
of the sign of the averaged Polyakov
loop on the topological susceptibility.

While these results (see Sect.~\ref{sec:topological_susceptibility})
could have been expected, the remaining observations concerning the spectral 
properties might come as a surprise, if not at a qualitative then at a 
quantitative level.

Assuming antiperiodic boundary conditions for the fermionic fields
throughout this paper 
we have discussed the behavior of the chiral condensate defined
via the Banks-Casher relation below and above the phase transition. 
A qualitatively different behavior of the spectral density has been 
observed at $T > T_c$ for ensembles having opposite sign of the Polyakov 
loop $L$ (see Sect.~\ref{sec:spectral_density}). 
The spectral density is decreasing with rising temperature for $L > 0$ 
until it is finally disrupting around $T \sim 1.5~T_c$. On the other hand, 
for $L < 0$ the density begins to rise with increasing temperature.

Although the spectral gap $g_\lambda$, defined as the average of the lowest 
non-zero eigenvalue, as a function of temperature shows qualitative
agreement with respective data for $SU(3)$ gauge theory (see Fig.~\ref{fig:fig5})
we observed that 
the respective volume dependence is completely different. 
We found that $g_\lambda$ tends fast to zero with increasing volume for 
$L > 0$ at $T/T_c=1.15$ and for $L < 0$ for both $T/T_c=1.15$ and 1.55 
in contrast to the situation in $SU(3)$ where a rather weak volume dependence 
of the gap was observed in both real and complex $L$ sectors.

Increasing of the gap $g_{\lambda}$ with increasing temperature for $L > 0$ 
and fixed lattice size is explained by the strong volume dependence of the 
number of near-zero modes as can be seen from Fig.~\ref{fig:fig16}. 
We are making measurements on the lattice, with $N_s=20$ defining its
physical size, which decreases with increasing temperature, resulting in a 
decreasing number of near-zero modes. The consequence is that not each lattice
configuration has these modes. This in turn makes the {\it ensemble average}
$g_{\lambda}$ to increase with temperature. 

The localization properties of the eigenmodes have been investigated 
in terms of the inverse participation ratio (IPR), and the dependence of the
average IPR on volume and discretization has been monitored all over the
lower part of the spectrum (see Sect.~\ref{sec:localization}. 
We repeat here only two remarkable facts. At first, sizeable lattice spacing 
effects have been found only for the zero modes (close to $T_c$) which is
pointing to a dimensionality $\approx 2$. Second, at $L < 0$, the eigenmodes 
responsible for the non-vanishing chiral condensate above $T_c$, 
that are filling the region up to $\lambda \sim 400 {\rm~MeV}$, are very 
delocalized compared to the isolated modes that exist in that spectral range 
for $L > 0$.

Special considerations have been devoted in Sect.~\ref{sec:Edwards} to the 
near-zero modes found in the sub-ensemble with $L > 0$. 
The spectral density has been divided into an interval with small eigenvalues 
and the rest with all larger eigenvalues, motivated by the observation 
that for large enough temperatures $T \approx 1.5~T_c$ these two parts of the 
spectrum become well separated, whereas a small number of near-zero modes 
still exists. The topological susceptibility (which had 
been defined through the total topological charge $Q$ obtained by counting 
the zero modes alone) now has been compared with the density of topological 
objects defined {\it including} the near-zero modes. 
Good agreement was found. This justifies the dilute-gas approximation, 
$\chi = \rho_{top}$ with charge $Q \approx \pm 1$ objects. 
This is corroborated by the observation that the multiplicity of all 
topological objects follows the Poisson distribution with reasonable 
accuracy as indicated by Fig.~\ref{fig:fig15} and was reported earlier 
in Ref.~\cite{Edwards}. Further, the percentage of near-zero modes among the 
lower part of the spectrum ({\it i.e.} including also the true zero-modes) 
has been measured for different temperatures and volumes, see Fig.~\ref{fig:fig16}. 
It was found to rise with increasing volume, and the apparent temperature 
suppression observed in our measurements might be mainly explained as a volume 
effect. 

Based on all these observations, emphasizing the special role of the near-zero 
modes, an alternative definition of the spectral gap $\bar{g}_\lambda$ 
{\it excluding} the near-zero modes has been proposed. 
Following this definition, a non-vanishing infinite-volume limit of 
$\bar{g}_{\lambda}$ in the deconfined phase {\it at all temperatures}
$T > T_c$ now seems possible for the $L > 0$ sub-ensemble, in clear distinction 
to the other definition and, as we have seen, to the $L < 0$ sub-ensemble.
Then we end up with an unusual situation in the temperature interval 
$T_c < T < 2~T_c$ : a finite density of the near-zero modes 
coexists with a gap between these modes and the bulk of the spectrum.  
Whether we really may speak about broken chiral symmetry depends on 
a nonvanishing infinite-volume limit of the spectral density 
$\rho_{sm}(\lambda)$ at $\lambda=0$. Otherwise we could ignore the 
presence of the near-zero modes for this question
and consider the new definition $\bar{g}_\lambda$ 
of the spectral gap as a proper one.
In any case, more studies of the thermodynamical limit are necessary
in order to confirm or reject the possibility to reconcile, in that limit,
the restoration of chiral symmetry, $\rho(\lambda=0)=0$, with a gap 
$g_{\lambda} \to 0$.~\footnote{We thank one of the anonymous referees for 
pointing out to us such a scenario.}

It shall be noted that by now, the ``kinematic'' understanding for the relation 
between chiral symmetry breaking (and restoration) and the sign of an eventually
nonzero value of the Polyakov loop (or, equivalently, on the change
of the boundary conditions in the Dirac operator) is much more developed
than at the time when our study was begun~\cite{Lattice2007}.

A series of papers~\cite{linking,complete,thin_and_dressed,dual,Synatschke:2008yt}
discusses the connection between the spectrum of
a (fairly general) Dirac operator and the breaking of center symmetry,
{\it i.e.} the emergence of a nonvanishing Polyakov loop.
The most concise quantity in this respect is the ``dual quark condensate''
defined in Ref.~\cite{dual}, the Fourier transform of the standard quark
condensate
$$\Sigma = \frac{1}{V} \langle {\rm Tr} \left[(m + D_{\phi})^{-1}\right] \rangle $$
with respect to the angle $\phi$ modifying the boundary condition of the
Dirac operator $D_{\phi}$.
A ``fat Polyakov loop'' formed out of singly wrapped pathes
is obtained as the lowest ($n=1$) Fourier component.
Higher Fourier components correspond to multiply wrapped Polyakov loops.
The condition for a nonvanishing ``fat'' Polyakov loop is a nonvanishing
periodic component $\propto \cos(\phi)$ in the fermionic spectral density,
in other words, a nontrivial response of the spectral density to the change
of the boundary conditions.

A microscopic explanation of the onset of such a sensitivity of the Dirac spectrum
to the type of boundary conditions in the context of a nonvanishing Polyakov
loop of either sign needs to be worked out. What comes to mind is the interplay
of holonomy and topology of dyons~\cite{calorons}. This difference is accompanied
by a different localization behavior of the lowest fermionic eigenmodes in the
two sectors.

\section*{Acknowledgements}

The work of ITEP group is partially supported by RFBR grants
RFBR 08-02-00661a and RFBR 07-02-00237a, by the grant 
RFBR 06-02-04010-NNIOa together with the DFG grant 436 RUS 113/739/2, 
and the grant for scientific schools NSh-679.2008.2.
A considerable part of the computations has been performed on the 
MVS 50K multiprocessor system at the Moscow Joint Supercomputer Center.

The work of E.-M.~I. was supported by DFG through the Forschergruppe
FOR 465 (Mu932/2). 
We thank Maxim Chernodub and Christof Gattringer for discussions
on an earlier version. E.-M~I. is grateful to the 
Karl-Franzens-Universit\"at Graz for the guest position he holds 
while the revised version is written up.


\vspace{1.0cm}

\begin{thebibliography}{99}
\label{sec:bibliography}

\bibitem{Stephanov}
M.~A.~Stephanov,
Phys.~Lett.~{\bf B~375} (1996) 249 [hep-lat/9601001].

\bibitem{Chandrasekharan1}
S.~Chandrasekharan and N.~Christ,
Nucl.~Phys.~Proc.~Suppl.~{\bf 47} (1996) 527 [hep-lat/9509095].

\bibitem{Chandrasekharan2}
S.~Chandrasekharan, Dong~Chen, N.~Christ, W.~Lee, R.~Mawhinney, and P.~Vranas,
Phys.~Rev.~Lett.~{\bf 82} (1999) 2463 [hep-lat/9807018].

\bibitem{Gattringer}
Ch.~Gattringer, P.~E.~L.~Rakow, A.~Sch\"afer, and W.~S\"oldner,
Phys.~Rev.~{\bf D~66} (2002) 054502 [hep-lat/0202009].

\bibitem{Lattice2007}
V.~G.~Bornyakov, E.~V.~Luschevskaya, S.~M.~Morozov, M.~I.~Polikarpov,
E.-M.~Ilgenfritz, and M.~M\"uller-Preussker,
PoS~LAT2007 (2007) 315 [arXiv:0710.2799 (hep-lat)].

\bibitem{Neuberger}
H.~Neuberger,
Phys.~Lett.~{\bf B~417} (1998) 141 [arXiv:hep-lat/9707022].

\bibitem{Laliena}
P.~Hasenfratz, V.~Laliena, and F.~Niedermayer,
Phys.~Lett.~{\bf B427} (1998) 125 [hep-lat/9801021].
			
\bibitem{Adams}
D.~H.~Adams,
J.~Math.~Phys.~{\bf 42} (2001) 5522 [hep-lat/0009026].

\bibitem{calorons}
V.~G.~Bornyakov, E.-M.~Ilgenfritz, B.~V.~Martemyanov, S.~M.~Morozov,
M.~M\"uller-Preussker, and A.~I.~Veselov,
Phys.~Rev.~{\bf D 76} (2007) 054505 [arXiv:0706.4206 (hep-lat)].

\bibitem{Niedermayer}
F.~Niedermayer,
Nucl.~Phys.~Proc.~Suppl.~{\bf 73} (1999) 105 [hep-lat/9810026].

\bibitem{Edwards}
R.~G.~Edwards, U.~M.~Heller, J.~E.~Kiskis, and R.~Narayanan,
Phys.~Rev.~{\bf D~61} (2000) 074504 [arXiv:hep-lat/9910041].

\bibitem{universality}
V.~G.~Bornyakov, E.-M.~Ilgenfritz, and M.~M\"uller-Preussker,
Phys.~Rev.~{\bf D 72} (2005) 054511 [hep-lat/0507021].

\bibitem{Engels}
J.~Engels, J.~Fingberg, and M.~Weber,
Nucl.~Phys.~{\bf B 332} (1990) 737.
							  
\bibitem{arpack}
ARPACK source: http://www.caam.rice.edu/software/ARPACK/

\bibitem{Pullirsch}
Ch.~Gattringer and R.~Pullirsch,
Phys.~Rev.~{\bf D~69} (2004) 094510 [hep-lat/0402008].

\bibitem{DiGiacomo}
B.~Alles, M.~D'Elia, and A.~Di~Giacomo,
Phys.~Lett.~{\bf B~412} (1997) 119 [hep-lat/9706016].

\bibitem{BanksCasher}
T.~Banks and A.~Casher,
Nucl.~Phys.~{\bf B~169} (1980) 103.

\bibitem{Ilgenfritz}
E.-M.~Ilgenfritz, K.~Koller, Y.~Koma, G.~Schierholz, T.~Streuer,
and V.~Weinberg,
Phys.~Rev.~{\bf D~76} (2007) 034506 [arXiv:0705.0018 (hep-lat)].

\bibitem{Gubarev}
F.~V.~Gubarev, S.~M.~Morozov, M.~I.~Polikarpov, and V.~I.~Zakharov,
JETP~Lett.~{\bf 82} (2005) 343, Pisma~Zh.~Eksp.~Teor.~Fiz.~{\bf 82} (2005) 381
[hep-lat/0505016].

\bibitem{Koma}
Y.~Koma, E.-M.~Ilgenfritz, K.~Koller, G.~Schierholz, T.~Streuer,
and V.~Weinberg,
PoS~LAT2005 (2006) 300, [hep-lat/0509164].

\bibitem{Weinberg2}
V.~Weinberg, E.-M.~Ilgenfritz, K.~Koller, Y.~Koma, G.~Schierholz,
and T.~Streuer,
PoS~LAT2006 (2006) 078, [hep-lat/0610087].

\bibitem{Weinberg1}
V.~Weinberg, E.-M.~Ilgenfritz, K.~Koller, Y.~Koma, G.~Schierholz,
and T.~Streuer,
PoS~LAT2005 (2006) 171, [hep-lat/0510056].

\bibitem{Weinberg3}
V.~Weinberg {\it et al.} (DIK Collaboration),
PoS~LAT2007 (2007) 236, [arXiv:0710.2565 (hep-lat)].

\bibitem{0107016}
Ch.~Gattringer, M.~G\"ockeler, P.E.L.~Rakow, St.~Schaefer, and A.~Sch\"afer,
Nucl.~Phys.~{\bf B 617} (2001) 101 [hep-lat/0107016].

\bibitem{0105023}
Ch.~Gattringer, M.~G\"ockeler, P.E.L.~Rakow, St.~Schaefer, and A.~Sch\"afer,
Nucl.~Phys.~{\bf B 618} (2001) 205 [hep-lat/0105023].

\bibitem{linking}
Ch.~Gattringer,
Phys.~Rev.~Lett.~{97} (2006) 032003 [hep-lat/0605018].

\bibitem{complete}
F.~Bruckmann, Ch.~Gattringer, and Ch.~Hagen,
Phys.~Lett.~{\bf B 647} (2007) 56 [hep-lat/0612020].

\bibitem{thin_and_dressed}
Ch.~Hagen, F.~Bruckmann, E.~Bilgici, and Ch.~Gattringer,
PoS~LAT2007 (2007) 289, [arXiv:0710.0294 (hep-lat)].

\bibitem{dual}
E.~Bilgici, F.~Bruckmann, Ch.~Gattringer, and Ch.~Hagen,
Phys.~Rev.~{\bf D 77} (2008) 094007 [arXiv:0801.4051 [hep-lat]].

\bibitem{Synatschke:2008yt}
F.~Synatschke, A.~Wipf, and K.~Langfeld,
Phys.~Rev.~{\bf D 77} (2008) 114018 [arXiv:0803.0271 [hep-lat]].
		
\end{thebibliography}
\end{document}